\documentclass[11pt,a4paper]{article}
\usepackage{jcappub}

\begin{document}

\title{Combined cosmological tests of a bivalent tachyonic dark energy
scalar field model}
\author{Zolt\'{a}n Keresztes and L\'{a}szl\'{o} \'{A}. Gergely}

\affiliation{Departments of Theoretical and Experimental Physics, University
of Szeged, D\'{o}m t\'{e}r 9, 6720 Szeged, Hungary}

\emailAdd{zkeresztes@titan.physx.u-szeged.hu,gergely@physx.u-szeged.hu}

\abstract{
A recently investigated tachyonic scalar field dark energy dominated
universe exhibits a bivalent future: depending on initial parameters can run either into a de Sitter
exponential expansion or into a traversable future soft singularity
followed by a contraction phase. We also include in the model (i) a tiny
amount of radiation, (ii) baryonic matter ($\Omega _{b}h^{2}=0.022161$,
where the Hubble constant is fixed as $h=0.706$) and (iii) cold dark matter
(CDM). Out of a variety of six types of evolutions arising in a more subtle
classification, we identify two in which in the past the scalar field
effectively degenerates into a dust (its pressure drops to an
insignificantly low negative value). These are the evolutions of type IIb
converging to de Sitter and type III hitting the future soft singularity. We
confront these background evolutions with various cosmological tests,
including the supernova type Ia Union 2.1 data, baryon acoustic oscillation
distance ratios, Hubble parameter-redshift relation and the cosmic microwave
background (CMB) acoustic scale. We determine a subset of the evolutions of
both types which at 1$\sigma $ confidence level are consistent with all of
these cosmological tests. At perturbative level we derive the CMB
temperature power spectrum to find the best agreement with the Planck data
for $\Omega _{CDM}=0.22$. The fit is as good as for the $\Lambda $CDM model
at high multipoles, but the power remains slightly overestimated at low
multipoles, for both types of evolutions. The rest of the CDM is effectively
generated by the tachyonic field, which in this sense acts as a combined
dark energy and dark matter model.}

\arxivnumber{1408.3736}

\maketitle

\section{Introduction}

The discovery of the accelerated expansion rate of the Universe at late
times \cite{acc} induced the necessity to model dark energy, the unknown
energy form responsible for such a phenomenon. Beyond the simple, but
conceptually unsatisfactory cosmological constant wide classes of dark
energy models were investigated. The most common models introduce a scalar
field $\phi $ as dark energy candidate. In the simplest, quintessence models 
\cite{quintessence} the dynamics of the scalar field is encompassed in the
Lagrangian density $\mathcal{L}=\sqrt{-g}L$ through the canonical Lagrangian 
$L=X-\mathcal{V}\left( \phi \right) $ (here $g,~X,~\mathcal{V}$ stand for
the metric determinant, standard kinetic term $X$ and potential term $%
\mathcal{V}$). Generalized k-essence models \cite{kessence} exhibit a
Lagrangian with non-standard dependence of the kinetic term, hence $%
L=P\left( \phi ,X\right) $, with $P$ an arbitrary function. A particular
subcase of the latter is obtained when $L=-V\left( \phi \right) \sqrt{1-2X}$%
, thus it has the Dirac-Born-Infeld form \cite{tachyon}. In this latter case
the scalar $\phi \equiv T$ is known as a tachyonic field.

If the scalar field depends solely on time (which is the case in the
presence of cosmological symmetries), its energy-momentum tensor
characterizes a perfect fluid. In particular, a time-dependent, homogeneous
tachyonic field can be perceived as a perfect fluid. (When the potential $V$
is a constant, this fluid becomes the Chaplygin gas, which together with its
generalizations was also studied as a dark energy candidate \cite{Chaplygin}%
.)

It was shown that tachyonic fields with inverse square law or exponential
potentials could play the role of dark energy, as they were found consistent
with type Ia Supernovae (SNIa) data and with the requirements of structure
formation \cite{tachobs1}. Furthermore the luminosity-redshift relation
arising from SNIa data, the baryon acoustic oscillation (BAO) distance
ratios from recent galaxy surveys (BAO are the imprint in the distribution
of matter of the sound horizon at the last scattering surface), the Hubble
constant measurement from Hubble Space Telescope data and the cosmic
microwave background (CMB) temperature anisotropy can be explained by a
scalar field (quintessence or tachyonic) dark energy with equation of state $%
p=c_{a}^{2}\rho +C$\ with constants $c_{a}^{2}$ and $C$ \cite{tachobs2}.

The dynamics of tachyonic cosmological models can be quite rich, depending
on the chosen potential. For the trigonometric potential discussed in Ref. 
\cite{tach0} some of the future evolutions, rather then asymptoting to the
de Sitter attractor, will exhibit a slowdown of the accelerated expansion
(during which the tachyonic field still behaves as dark energy), then
continue through a decelerated regime (when the tachyonic field ceases to
mimic dark energy and it evolves superluminally) until the deceleration
reaches infinite value and the expansion suddenly stops. This is a specific
example of a future sudden singularity \cite{sudden} dubbed Big Brake \cite%
{tach0}, characterized by finite values of the scale factor, vanishing
energy density and Hubble parameter, but diverging deceleration and infinite
pressure.

The question naturally arises whether such evolutions can actually be
realized in our Universe. In Ref. \cite{tach1} the observational data on
SNIa was confronted with the evolutions of the universe filled with such a
one-parameter family of tachyonic models. Among the set of the trajectories
of the model compatible with the SNIa data at 1$\sigma $ level, a subset was
found to evolve into a Big Brake. The time scales for reaching this
singularity are finite, at the order of the present age of the universe.

As shown in Ref. \cite{tach2} the infinities appearing at the Big Brake only
affect the geodesic deviation equation, in the form of infinite tidal
forces. The geodesics themselves remain regular, hence they can be continued
through the singularity. Once matter particles have passed through, they
will determine the new geometry, which turned out to be a recollapsing one,
eventually reaching a Big Crunch.

The SNIa test works well also when baryonic and cold dark matter (CDM) are
added to the system \cite{tach3}. The combination of the tachyonic scalar
field and dust however leads to an additional problem when reaching the
singularity. Despite the tachyonic energy density vanishing at the
singularity, the dust still arrives with a nonzero energy density there,
hence the expansion rate is nonvanishing. Similar features emerge when
adding a dust component to an anti-Chaplygin gas. In both cases the Hubble
parameter acquires a nonzero value at the singularity due to the dust
component, implying further expansion. With continued expansion however,
both the energy density and the pressure would become ill-defined, hence
only a contraction would be allowed. The paradox is resolved by suitably
redefining the anti-Chaplygin gas in a distributional sense \cite{chap}.
Then due to a sudden reversal of the expansion rate (a jump in the Hubble
parameter) a contraction could instantly follow the expansion phase. This is
analogous to a ball bouncing back in a perfectly elastic manner from a wall.

As an alternative, certain transformations of the properties of both the
anti-Chaplygin gas and the tachyonic scalar field could lead to a smooth
passage through the soft singularity even in the presence of a dust
component. The expansion is continued for a while after the singularity,
with a full stop arising later on, followed by a contraction, a second
passage through the singularity and then further contraction until the Big
Crunch is reached \cite{tach3}. By analogy this process is similar to
modeling the deformations of the ball during the collision process with the
wall, which will lead to a full stop of the ball at the detriment of its
temporary deformation.

A distinct question is how the tachyonic scalar field model evolved in the
past. A purely theoretical study \cite{tach0} indicated that there are five
types of cosmological evolutions, all emerging from a Big\ Bang type
singularity (see Fig. \ref{Fig1}). Along type III trajectories the tachyonic
field exhibits negative pressure in the first era of the evolutions
(including a region of the velocity phase diagram where it can mimic dark
energy), however the pressure becomes positive later on and the field
evolves into a Big Brake singularity. By contrast, in the evolutions of type
V the tachyonic field exhibits positive pressure all the time (hence it
doesn't have a dark energy regime, proving itself incompatible with the
present day acceleration) and evolves into a Big Brake. In the evolutions of
type I and IV both regimes are present: these evolutions all start with a
positive pressure regime (hence superluminal evolution of the tachyonic
field), then the pressure turns negative (and the tachyonic field evolves
subluminally) so that in principle they can mimic dark energy. The evolution
of type I goes into the de Sitter attractor, while the evolution IV allows
for another change of the sign of the pressure and finally run into a Big
Brake. The type II trajectories arise from the Big Bang at $s^{2}=1$ and end
in the de Sitter attractor.

In this work we focus on the past evolutions, by confronting them with a
powerful set of cosmological observations. Our aim is to find the evolutions
which could be realized in our Universe.

In section \ref{PureTachyon} we briefly present the tachyonic scalar field
model with trigonometric potential (for simplicity we do not include other
matter types in this section) and revisit the compatibility with the SNIa
observations based on the most recent available Union 2.1 data set \cite%
{Union21}. We prove here for the first time that only the evolutions of type
I, II and III are compatible with SNIa data at 1$\sigma $\ confidence level,
disruling those of type IV, which on purely theoretical grounds were also
allowed. A further analysis based on test with SNIa and Hubble parameter
data \cite{H1}, \cite{H2} shows that only the types II and III are allowed
at 1$\sigma $\ confidence level. Next we prove that the evolution of the
effective equation of state parameter disrules the trajectories of type I
and a subclass of type II evolutions denoted IIa, as they built up
significant pressure in the distant past. They also fail to obey basic
stability requirements, as the square of the speed of sound becomes
negative. The rest of the trajectories of type II denoted IIb and all of
type III survive these tests. The division of the trajectories of type II
into IIa and IIb enriches the phase diagram, which now contains six types of
evolutions.

In section \ref{Tests} we proceed with the analysis of a more realistic
universe, which includes radiation, baryonic matter and CDM. Further tests
of the trajectories of type IIb and III are performed. In this setup we
identify the initial (present) values for the tachyonic parameters
characterizing the trajectories selected by SNIa data at 1$\sigma $
confidence level. Then we achieve subsequent substantial reductions of this
parameter region by successive inclusions of constraints from BAO, from the
Hubble parameter data and from CMB acoustic scale. All these constraints
refer to the cosmological evolution at background level.

In section \ref{Perturbations} we develop a perturbative description at the
linear level of the tachyonic scalar field, which is a prerequisite in
deriving the CMB temperature power spectrum, also presented there. In the
process the amount of CDM required in the tachyonic universe is found.
Section \ref{CR} contains the concluding remarks.

We employ the system of units $c=1$ and $8\pi G/3=1$. Throughout the paper
the tachyonic parameter is fixed as $\mathsf{k}=0.44$ and the present value
of the Hubble parameter at $H_{0}=70.6$ km/sec/Mpc \cite{Riess}-\cite%
{Delubac}.

\section{Background evolution of the flat Friedmann universe filled with
tachyonic scalar field \label{PureTachyon}}

In this section we present the background evolution of the universe
dominated by a tachyonic scalar field with a special trigonometric potential.

\subsection{Background dynamics and velocity phase diagram}

We consider a flat Friedmann universe%
\begin{equation}
ds^{2}=dt^{2}-a^{2}\left( t\right) \sum_{\alpha }\left( dx^{\alpha }\right)
^{2}~,
\end{equation}%
(with $x^{\alpha }$, $\alpha =1,2,3$ the Cartesian coordinates and $a$ the
scale factor). The dynamics is governed by the Raychaudhuri (second
Friedmann) equation%
\begin{equation}
\dot{H}=-\frac{3}{2}\left( \rho +p\right) ~  \label{Raych}
\end{equation}%
and the continuity equation 
\begin{equation}
\dot{\rho}+3H\left( \rho +p\right) =0~.  \label{cont}
\end{equation}%
(Here $H\equiv \dot{a}/a$ is the Hubble parameter, $\rho $ the energy
density and $p$ the pressure of the ideal fluid filling the universe, while
a dot denotes derivatives with respect to the cosmological time $t$.) The
(first) Friedmann equation 
\begin{equation}
H^{2}=\rho  \label{Friedmann}
\end{equation}%
stands as a first integral of the system (\ref{Raych})-(\ref{cont}).

The tachyonic Lagrangian is given by \cite{Sen} 
\begin{equation}
L=-V(T)\sqrt{1-g^{ij}\left( \partial _{i}T\right) \left( \partial
_{j}T\right) }~,  \label{L}
\end{equation}%
where $V(T)$ is a potential. A spatially homogeneous scalar field $T(t)$
evolves according to%
\begin{equation}
\frac{\dot{s}}{1-s^{2}}+3Hs+\frac{V_{,T}}{V}=0~,  \label{KG}
\end{equation}%
where $s=\dot{T}$ and $,T$ denotes the partial derivative with respect to $T$%
.

The energy-momentum tensor $T_{ab}$ can be obtained from the variation of
the action for tachyonic field with respect to the metric, and it can be
decomposed with respect to an observer with 4-velocity $u^{a}$ as 
\begin{equation}
T_{ab}=\rho u_{a}u_{b}+2q_{(a}u_{b)}-ph_{ab}+\pi _{ab}\ .  \label{EMTdec}
\end{equation}%
Here $\rho $, $q_{a}$, $p$ and $\pi _{ab}$ are the energy density, the
energy current 3-vector, the isotropic pressure and the symmetric,
trace-free, anisotropic pressure 3-tensor of the matter. With the choice $%
u_{a}=\left( dt\right) _{a}$ a spatially homogeneous tachyonic field becomes
an ideal fluid ($q_{a}=0$, $\pi _{ab}=0$) with energy density 
\begin{equation}
\rho ^{\left( T\right) }=\frac{V(T)}{\sqrt{1-s^{2}}}~,  \label{rho}
\end{equation}%
and pressure 
\begin{equation}
p^{\left( T\right) }=-V(T)\sqrt{1-s^{2}}~.  \label{p}
\end{equation}%
As long as the potential is real, the Lagrangian density, $\rho ^{\left(
T\right) }$ and $p^{\left( T\right) }$ are well defined only for $s^{2}\leq
1 $. Outside this range the energy density and pressure remain well defined
for an imaginary potential. Note that the fluid becomes effectively
barotropic with the equation of state parameter $w_{T}=p^{\left( T\right)
}/\rho ^{\left( T\right) }=s^{2}-1$. Hence for subluminal ($s^{2}<1$)
tachyonic field evolutions $w_{T}<0$ and (for a positive potential) the
pressure is negative, allowing in principle for violations of the strong
energy condition, rendering the tachyonic field into the dark energy regime.
At $s^{2}=1$ the pressure vanishes, the fluid becoming dust.

We are interested in the dynamics generated by the simple trigonometric
potential \cite{tach0}:%
\begin{equation}
V(T)=\frac{\Lambda \sqrt{1-(1+\mathsf{k})y^{2}}}{1-y^{2}}~,  \label{pot}
\end{equation}%
where%
\begin{equation}
y=\cos \left[ \frac{3}{2}{\sqrt{\Lambda \,(1+\mathsf{k})}\,T}\right] ~
\label{pot1}
\end{equation}%
is an alternative scalar field variable, while $\Lambda >0$ and $-1<\mathsf{k%
}<1$ are the two parameters of the model. The system is invariant under the
simultaneous parity changes%
\begin{equation}
y\rightarrow -y\ ,\ \ s\rightarrow -s\ ,  \label{sym}
\end{equation}%
which generates a double coverage of the dynamics of such a tachyon-filled
universe in these velocity phase-space variables.

For numerical investigations it is worth to introduce the following
dimensionless quantities \cite{tach1}:%
\begin{equation}
\hat{H}=\frac{H}{H_{0}},\,\hat{V}=\frac{V}{H_{0}^{2}},\,\Omega _{\Lambda }=%
\frac{\Lambda }{H_{0}^{2}},\,\hat{T}=H_{0}T~,  \label{new-var}
\end{equation}%
and the redshift $z$ as independent variables. Then the equations of motion
become%
\begin{equation}
\hat{H}^{2}=\frac{\hat{V}}{\sqrt{1-s^{2}}}\ ,  \label{Hhat}
\end{equation}%
\begin{equation}
\frac{dy}{dz}=\frac{3\sqrt{\ \Omega _{\Lambda }\left( 1+\mathsf{k}\right)
(1-y^{2})}}{2\left( 1+z\right) \hat{H}}s\ ,  \label{yEq}
\end{equation}%
\begin{equation}
\frac{1+z}{1-s^{2}}\frac{ds}{dz}=3s+\frac{\hat{V}_{,\hat{T}}}{\hat{H}\hat{V}}%
~,  \label{sEq}
\end{equation}%
where $\hat{V}$ and $\hat{V}_{,\hat{T}}/\hat{V}$ are given by 
\begin{equation}
\hat{V}=\frac{\ \Omega _{\Lambda }\sqrt{1-\left( 1+\mathsf{k}\right) y^{2}}}{%
1-y^{2}}\ ,  \label{VT}
\end{equation}%
\begin{equation}
\frac{\hat{V}_{,\hat{T}}}{\hat{V}}=\frac{3\sqrt{\ \Omega _{\Lambda }\left( 1+%
\mathsf{k}\right) }\left[ \mathsf{k}-1+\left( 1+\mathsf{k}\right) y^{2}%
\right] }{2\sqrt{1-y^{2}}\left[ 1-\left( 1+\mathsf{k}\right) y^{2}\right] }%
y\ .
\end{equation}%
Since $\hat{H}\left( z=0\right) =1$, the first integral (\ref{Hhat}) gives a
relation between the parameters: $\mathsf{k}$, $\Omega _{\Lambda }$, $%
y\left( z=0\right) =y_{0}$ and $s\left( z=0\right) =s_{0}$. The latter two
parameters fix the initial conditions for the tachyonic scalar field.
Remarkably, the equations (\ref{yEq}) and (\ref{sEq}) do not depend on $%
\Omega _{\Lambda }$, as $\hat{H}\propto \sqrt{\Omega _{\Lambda }}$ and $\hat{%
V}_{,\hat{T}}/\hat{V}\propto \sqrt{\Omega _{\Lambda }}$ both hold. Therefore
the diagram showing the evolutions in the ($\sqrt{\Omega _{\Lambda }}\hat{T}=%
\sqrt{\Lambda }T$, $s$) or equivalently in the ($y$, $s$) planes for a given 
$\mathsf{k}$ and $\Omega _{\Lambda }$ will not dependent on the particular
chosen value of $\Omega _{\Lambda }$.

\begin{figure}[th]
\begin{center}
\vskip-2.5cm \includegraphics[height=14cm,angle=0]{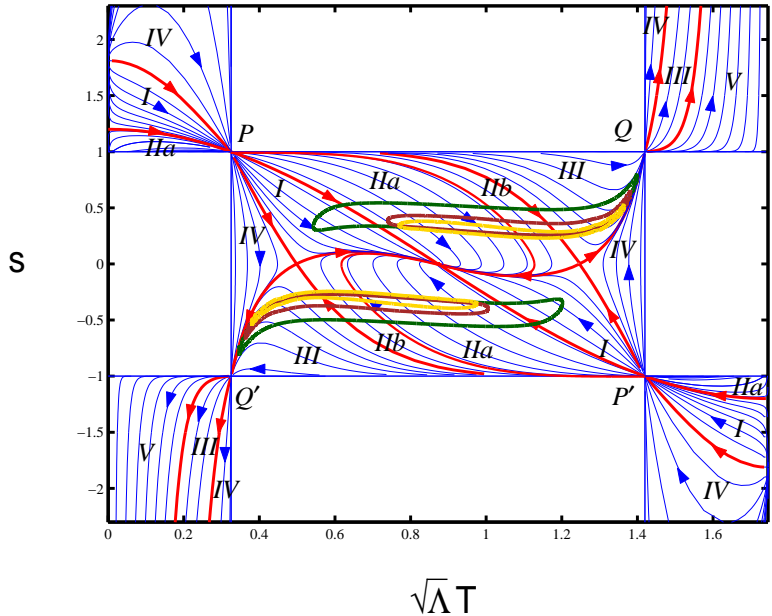} \vskip-2.5cm
\end{center}
\caption{The velocity phase diagram of the tachyonic scalar field dominated
universe, for a trigonometric potential and parameter $\mathsf{k}=0.44$.
Compared to the early version of this phase diagram \protect\cite{tach0}, 
\protect\cite{tach1} a new separatrix between IIa and IIb was added. The
richness of the dynamics is encoded in the six types of distinct tachyonic
evolutions. The de Sitter attractor is in the center of the figure while the
soft singularities arise in the vertical stripes on both the upper right and
the lower left corners. Two copies of all evolutions are consequence of the
symmetry (\protect\ref{sym}). Some of the evolutions cross the regions
compatible at 1$\protect\sigma $\ ($68.3\%$) confidence level with i) SNIa
data (delimited by a green contour), with ii) Hubble parameter data (yellow)
and with iii) SNIa+Hubble parameter data (brown).}
\label{Fig1}
\end{figure}

As shown on Fig. \ref{Fig1} the dynamics is quite rich. The attractive fix
point in the center of the figure (corresponding to $y=0$ and $s=0$)
represents the de Sitter evolution. Two types (I and II) of trajectories end
in this de Sitter attractor, but they originate in different Big Bang
singularities on the diagram. The evolutions of type I and the separatrices
between the trajectories of types I and II start from the points ($y=\pm
1,s=\pm \sqrt{1+1/\mathsf{k}}$), while type II from the lines $s^{2}=1$ \cite%
{tach0}. For most of the trajectories arriving to any of the four corners
(P, P', Q, Q') of the velocity phase diagram passage through the corners is
allowed \cite{tach0}, as these are the only points on the horizontal lines ($%
s^{2}=1$), which do not represent singularities (the vanishing of the
potential at $y=\pm (1+k)^{-1/2}$ assures that $s=\pm 1~$does not imply an
infinite energy density there). Nevertheless there is an exceptional
trajectory for each corner point (given by a vanishing integration constant $%
B$\ in Eqs. (78) and (82) of \cite{tach0}), which encounters a space-time
singularity at the respective corner point. Inside the central rectangle the
pressure is negative. In the side strips the pressure is positive, thus the
expansion of the universe is slowing down in those regimes. There the field
represents a pseudo-tachyon and has well-defined (real) Lagrangian, energy
density and positive pressure. Since in the process of evolving through the
corners (forward in time at Q, Q' and backward in time at P, P') to the side
strips both the potential $V$ and $\sqrt{1-s^{2}}$ become imaginary, a
redefined real potential $W(T)=iV(T)$ and $\sqrt{1-s^{2}}=i\sqrt{s^{2}-1}$
will be used in the Lagrangian\footnote{%
When $W\left( T\right) $ is a constant, the pseudo-tachyon field degenerates
into an anti-Chaplygin gas.}, which then becomes $W(T)\sqrt{s^{2}-1}$. In
this regime eventually a new type of soft cosmological singularity, the Big
Brake is reached by the trajectories of type III, IV and V, at $y\rightarrow
y_{BB}$ and $s^{2}\rightarrow \infty $ \cite{tach0}, \cite{tach1}-\cite%
{tach3}.\ From among them two types (III and IV) also exhibit an evolution
regime where the field has negative pressure. The trajectories of type III
are again born on the lines $s^{2}=1$ with $p^{\left( T\right) }<0$. The
curves of type IV originate at the same points as those of type I, then they
follows subsequent regimes with $p^{\left( T\right) }>0$, then $p^{\left(
T\right) }<0$ and again $p^{\left( T\right) }>0$, finally running into the
Big Brake singularity. The separatrices between trajectories of type I and
IV run into the unstable fix points ($y=\pm \sqrt{\left( 1-\mathsf{k}\right)
/\left( 1+\mathsf{k}\right) },s=0$) of the phase-velocity space \cite{tach0}%
. The separatrices between the evolutions of type II and III, originating on
the lines $s^{2}=1$ with $p^{\left( T\right) }<0$ run into the same unstable
fix points. From near the unstable fix points the trajectories either run
into the de Sitter attractor or into a Big Brake singularity generating
further separatrices between evolutions of type I and II or III and IV. The
separatrices between the trajectories of type III and V have $p^{\left(
T\right) }>0$ and originate in Big Bang singularities at the corner points
Q, Q', respectively. The trajectory of type V always has positive pressure.

On earlier versions of the velocity phase diagram (Fig. \ref{Fig1}),
discussed in Refs. \cite{tach0}, \cite{tach1}-\cite{tach3} it was not clear
whether the separatrix between the evolutions of type II and III reaches the
corner point P (P'). A thorough numerical investigation of the evolutions
this time made it possible to\ answer this question. We confirmed that some
of the evolutions of type II originate in the Big Bang type singularity
lying outside the central rectangle, hence they evolve through positive
pressures before they reach the corner point to pass in the rectangle region
with negative pressure. These trajectories, denoted IIa on the velocity
phase diagram however are complemented by other evolutions of type II, born
in a Big Bang singularity lying on the horizontal boundary of the rectangle.
Such trajectories, denoted IIb exhibit negative pressures throughout their
evolution. The velocity phase diagram Fig. \ref{Fig1} includes now new
separatrices between the trajectories of type IIa and IIb which originate in
Big Bang singularities at the corner points P, P', respectively, inside the
rectangle and both run into the de Sitter attractor. On the earlier version
of the diagram it was also not clear how the diagram depends on the
parameter $\Omega _{\Lambda }$ (or equivalently on $\Lambda $) which was
fixed. We have clarified this by giving the diagram in variables independent
of the actual value of $\Omega _{\Lambda }$.

\subsection{Confrontation with Supernovae Ia and Hubble parameter data \label%
{Test1}}

The cosmological test employing the supernovae data rely on the luminosity
distance ($d_{L}$)-redshift relation. In a flat Friedmann universe the
dimensionless luminosity distance $\hat{d}_{L}=H_{0}d_{L}$ satisfies the
relation 
\begin{equation}
\left( \frac{\hat{d}_{L}}{1+z}\right) ^{\prime }=\frac{1}{\hat{H}}~.
\label{dLz}
\end{equation}%
The confrontation of the tachyonic model with the Union 2.1 SNIa data set 
\cite{Union21} is done through a $\chi ^{2}$-test, repeating the procedure
of Ref. \cite{tach1}. In this paper we also perform a $\chi ^{2}$-test with
the Hubble parameter-redshift relation by computing%
\begin{equation}
\chi _{H}^{2}=\sum_{i=1}^{30}\frac{\left[ H_{th}\left( z_{i}\right) \!-\!%
\overline{H}_{obs}\left( z_{i}\right) \right] ^{2}}{\sigma _{i}^{2}}~.
\label{chiH}
\end{equation}%
Here $H_{th}\left( z_{i}\right) $\ and $\overline{H}_{obs}\left(
z_{i}\right) $\ are the values of the Hubble parameter at redshifts $z_{i}$\
predicted by the cosmological model and determined from the observations,
respectively, while $\sigma _{i}$\ is the scattering in $\overline{H}%
_{obs}\left( z_{i}\right) $. The data set on the Hubble parameter-redshift
relation was given in Refs. \cite{H1} and \cite{H2}. Recently a subset of
this data set was used to emphasize a tension with the $\Lambda $CDM model
(by computing the two-point $Omh^{2}$\ function) \cite{SSS}. Finally, we
perform a test with the combined SNIa and Hubble parameter data set by
calculating $\chi _{SNIa+H}^{2}=\chi _{SNIa}^{2}+\chi _{H}^{2}$, where $\chi
_{SNIa}^{2}$\ is the $\chi ^{2}$-value from the confrontation with SNIa data
set.

The first integral (\ref{Hhat}) evaluated at $z=0$ gives $\Omega _{\Lambda }$
(or equivalently $\Lambda $) as function of $y_{0}$, $s_{0}$. The confidence
level contours resulted from the $\chi ^{2}$-tests are represented on the
same velocity phase diagram (Fig. \ref{Fig1}), which is independent of $%
\Omega _{\Lambda }$. The SNIa test was not confronted with the velocity
phase diagram in previous analyses. We found that the evolutions compatible
with SNIa data at the 1$\sigma $\ confidence level are of the types I, II,
and III only. The trajectories of type IV, which in principle could have
allowed for accelerated expansion in recent times are disruled by SNIa data.
The fact that the trajectories of type V could not produce accelerated
expansion was obvious even without the SNIa test, as they do not venture
into the rectangle region with negative pressure. The inclusion of the test
with Hubble parameter-redshift relation shows that only the evolutions of
types II and III fall within the 1$\sigma $\ confidence level.

\subsection{Evolutions I and IIa disruled by nucleosynthesis and stability
arguments}

\begin{figure}[th]
\includegraphics[height=7cm,angle=270]{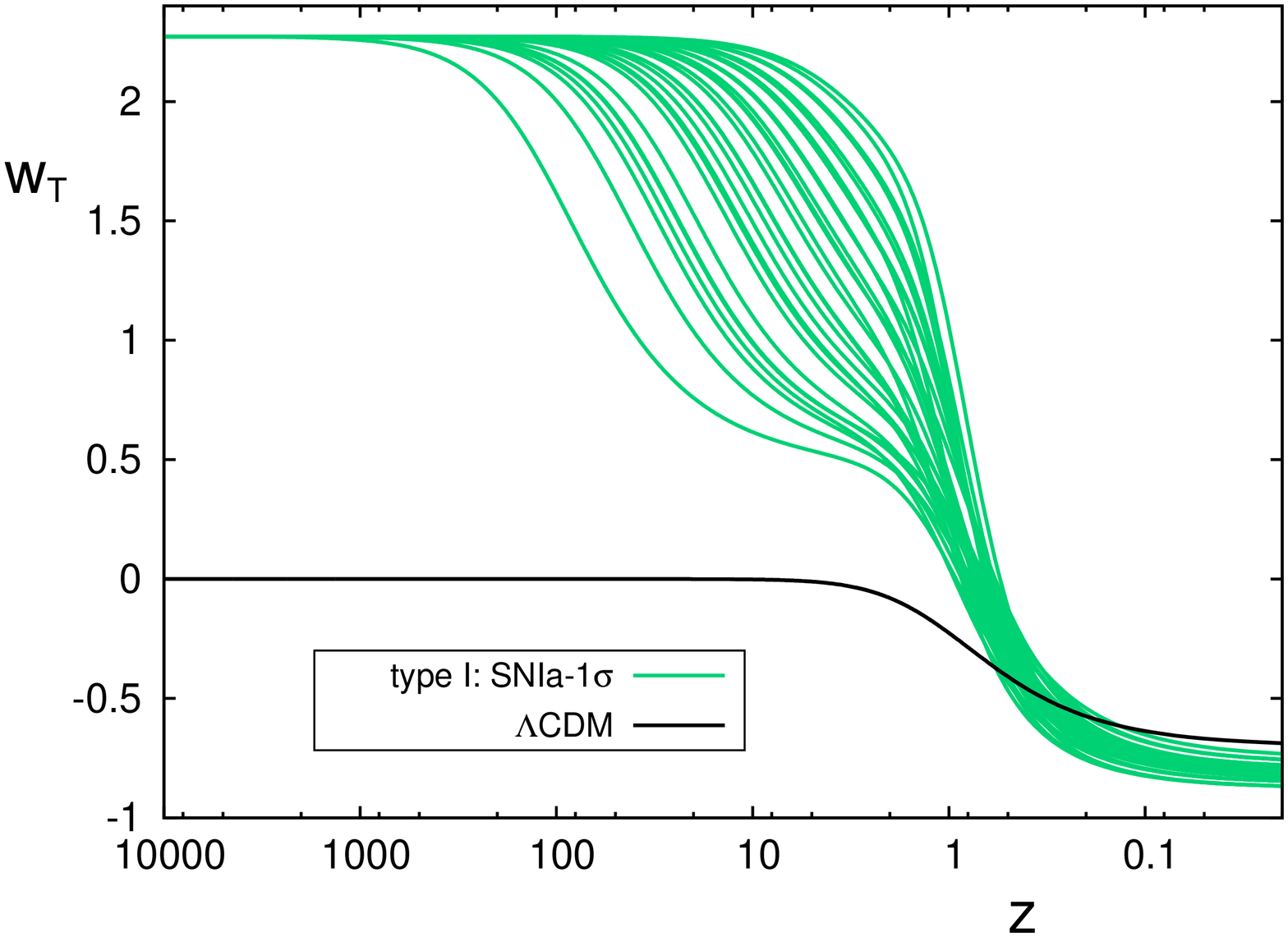} \hskip0.4cm%
\includegraphics[height=7cm,angle=270]{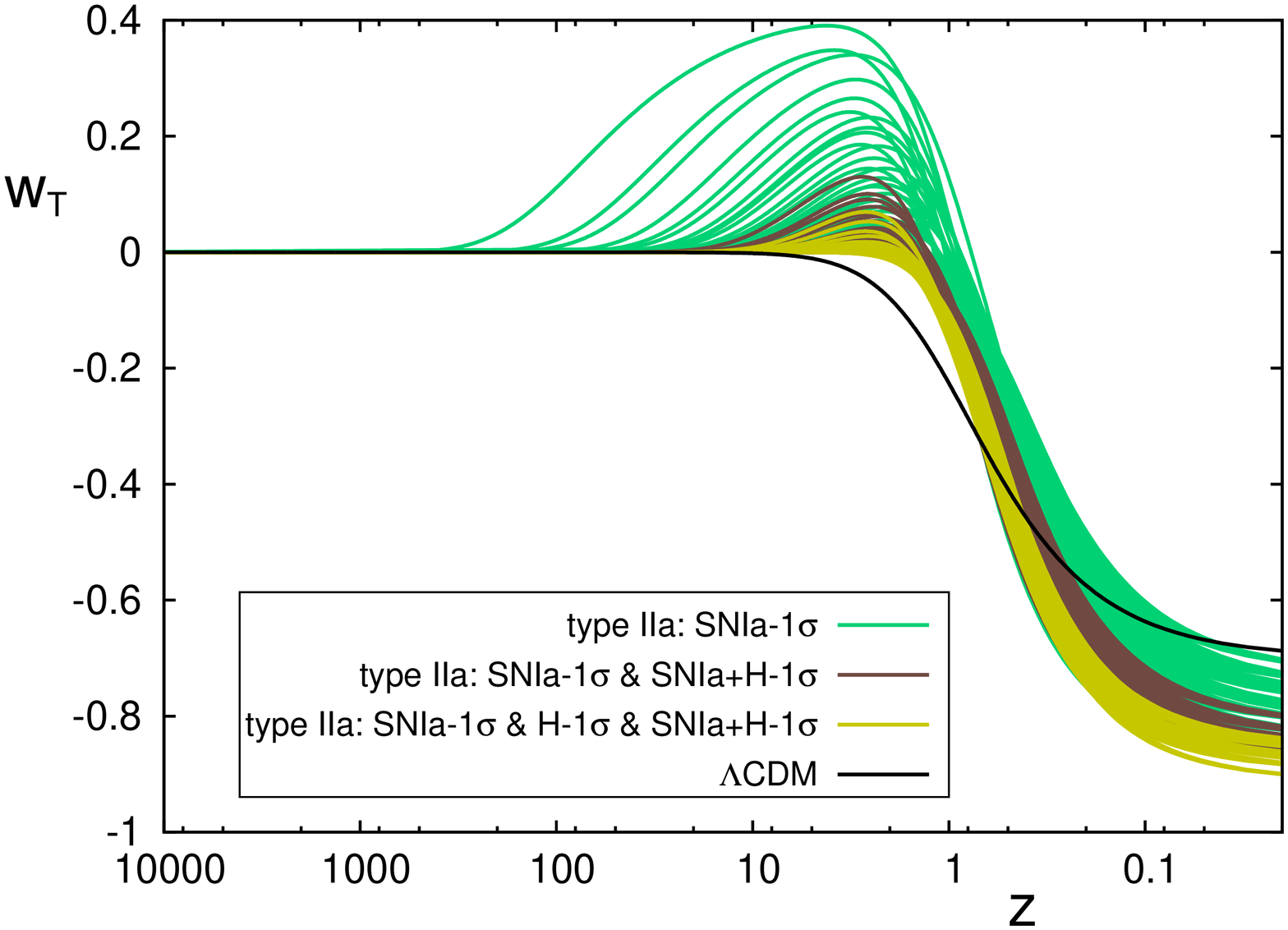} \vskip0.4cm %
\includegraphics[height=7cm,angle=270]{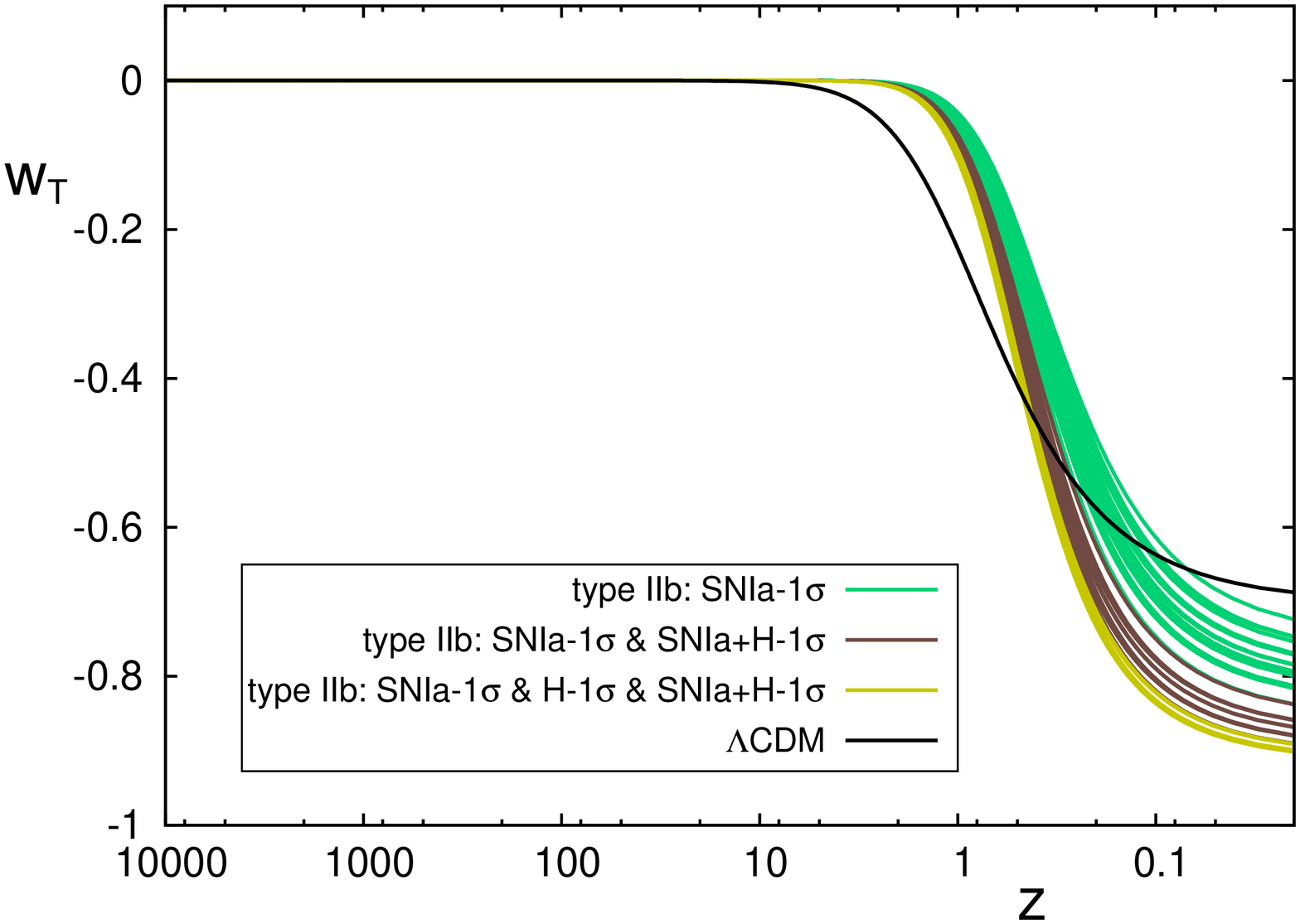} \hskip0.4cm%
\includegraphics[height=7cm,angle=270]{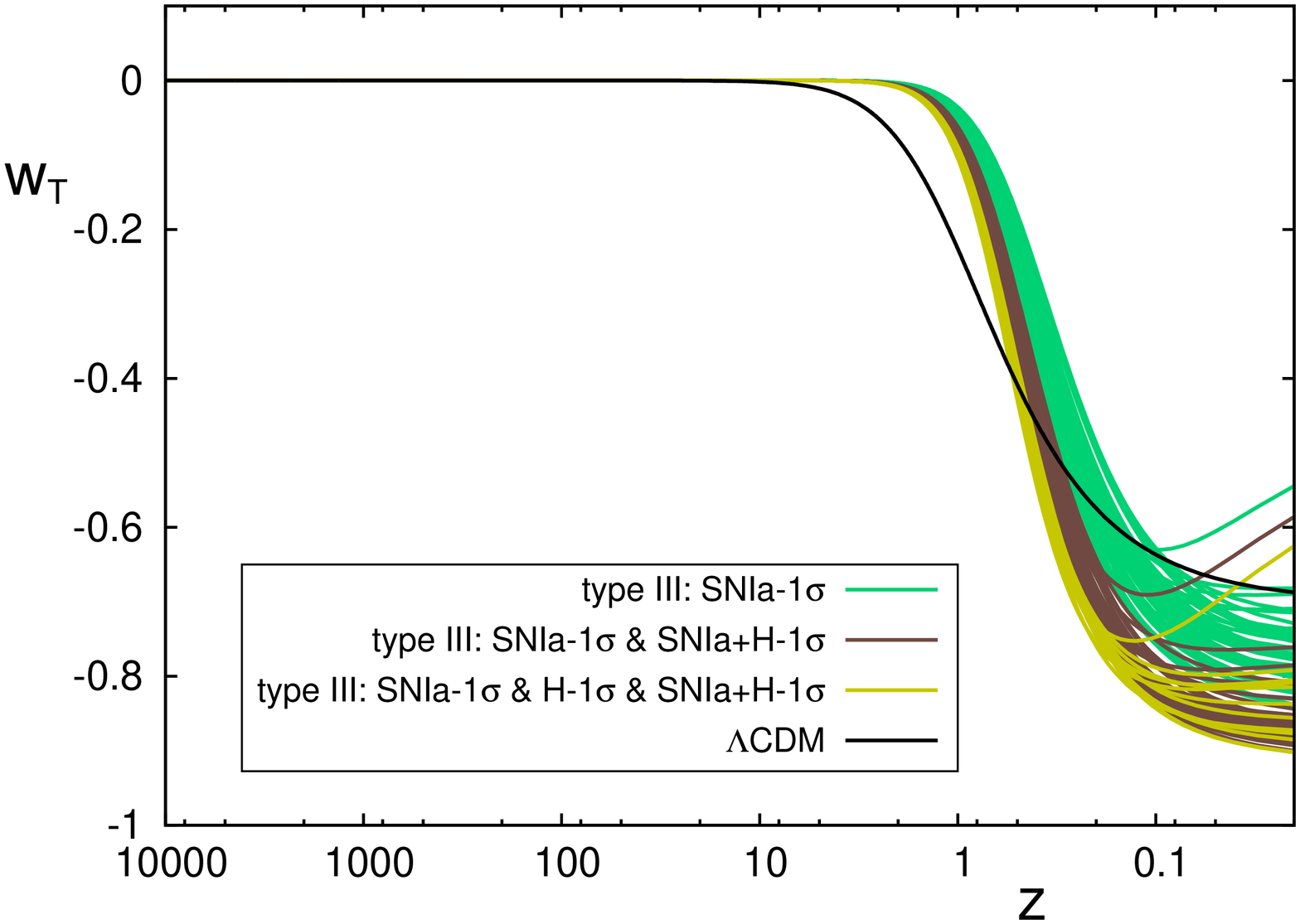} \vskip0.4cm
\caption{The distant past evolutions for different types of trajectories
(type I on the upper left panel, type IIa on the upper right panel, type IIb
on the lower left panel and type III on the lower right panel) fitting
within 1$\protect\sigma $\ confidence level with the Union 2.1 data set
(green curves), with Hubble parameter data (yellow curves) and with
SNIa+Hubble parameter data (brown curves). For comparison the evolution
generated in the framework of the flat $\Lambda $CDM model with $\Omega
_{\Lambda }=0.7$ and $\Omega _{m}=0.3$ is also shown. }
\label{Fig2}
\end{figure}

All evolutions surviving the SNIa test emerge from Big Bang like
singularities, these however are different for the trajectories of type I,
IIa, IIb and III. The past evolutions of these trajectories are depicted on
Fig. \ref{Fig2}. The trajectories compatible with the SNIa data, Hubble data
and SNIa+Hubble data at 1$\sigma $\ confidence levels are represented by
green, yellow and brown curves, respectively.


The evolutions of type I emerge from the singular point $y=\pm 1$\textbf{, }$%
s=\pm \sqrt{1+1/\mathsf{k}}$\textbf{\ }\cite{tach0}. As shown on the upper
right panel of Fig. \ref{Fig2}, with increasing redshift the barotropic
index $w_{T}=p^{\left( T\right) }/\rho ^{\left( T\right) }=s^{2}-1$
increases monotonically and converges to $\mathsf{k}^{-1}\approx 2.273$,
which is much larger as compared to the barotropic index of radiation. These
trajectories then could not be consistent with Big Bang Nucleosynthesis
(BBN), which stops when the plasma filling the Universe becomes dilute
enough to reduce the number of collisions among nuclei and cooled down
enough to stop the nuclei containing protons overcoming their electrostatic
repulsion. This approximately happens at $T\approx 0.1$ MeV, which in the $%
\Lambda $CDM model corresponds to the redshift $z\approx 4\times 10^{8}$.
Due to the high pressure however BBN is longer in the tachyonic model of
type I than in the $\Lambda $CDM model. In another line of reasoning, for
high value of the barotropic index\textbf{,} the continuity equation yields 
\begin{equation}
\rho _{T}\propto a^{-\frac{3}{k}\left( k+1\right) }\approx a^{-4.32}~,
\end{equation}%
implying higher energy density of the tachyonic field close to the Big Bang,
than for radiation, while the scale factor evolves as

\begin{equation}
a\propto t^{\frac{2k}{3\left( k+1\right) }}\approx t^{0.204}~,
\label{dSIearlyEvo}
\end{equation}%
exhibiting a much slower expansion than in either a dust or a radiation
dominated universe (where $a\propto t^{2/3}$ and $a\propto t^{1/2}$,
respectively). All of these suggest that by the end of the BBN the evolution
(\ref{dSIearlyEvo}) would have resulted in a higher ratio of the nuclei with
large mass numbers compared to Hydrogen as in a radiation dominated
universe. As the predictions of an early radiation dominated universe are
consistent with observations of the abundances of primordial light elements
(D, $^{3}$He, $^{4}$He, $^{7}$Li) \cite{nucleosynt}\textbf{, }type I
trajectories can be considered disruled.

Similar considerations disrule those evolutions of type IIa which run very
close to the separatrix between the evolutions of type I and IIa, as they
also build up large pressures (see Figs. \ref{Fig1}). For these evolutions,
once the universe passes the corner points, the pressure starts to increase
again, driving them away from the dust-dominated evolutions. The evolutions
presenting such pressure humps are however significantly disruled by the
combined SNIa test and Hubble parameter data.

Another aspect to comment on would be that outside the rectangle, inside the
stripes the pseudo-tachyonic field has a negative speed of sound squared.
Indeed, the pressure is growing when the energy density is decreasing,
hence, the derivative of the pressure with respect to energy density is
negative. The presence of an imaginary sound velocity means that the second
order equation governing the evolution of the perturbations instead of
oscillatory solutions exhibits two solutions with real exponents, one of
them positive, the other negative \cite{instab}. The positive one
corresponds to an exponentially growing mode, a Laplacian instability in the
evolution of the perturbations. Hence we disrule the models allowing for
such instabilities in the past. Note that the very same argument disrules
once again the evolutions of type I.

By contrast the trajectories of type IIb and III allow for a dark matter
dominated past ($w_{T}\approx 0$, see Fig. \ref{Fig2}), as they asymptote to
the singular horizontal lines of the velocity phase diagram Fig. \ref{Fig1}
and they never get away from there once they approach it. With a (today
insignificant) radiation component added, at the background level these
trajectories could be consistent with the early evolution of the Universe
(with radiation dominating at high redshift) and there are no instabilities
arise in the past either.

\section{Cosmological tests of an enhanced tachyonic universe at the
background level \label{Tests}}

In order to confront with various cosmological observations we need to make the
model more realistic. In the following subsection we introduce such an enhanced
model, while in the second subsection we perform a series of cosmological
tests available at background level, e.g. without working out the
perturbation formalism.

\subsection{Tachyonic universe encompassing radiation, baryons and CDM}

Starting from this subsection we include radiation, baryonic matter and CDM
in the model. In the flat $\Lambda $CDM model a detailed analysis of
temperature power spectrum of the cosmic microwave background shows that the
locations and the heights of the acoustic peaks are sensitive to $\Omega
_{b}h^{2}$ \cite{Mukhanov}. We fix the baryonic matter contribution as $%
\Omega _{b}h^{2}=0.022161$ cf. the Planck collaboration (taken from the last
column of Table 5 of Ref. \cite{PlanckXVI}). In the late universe the energy
density of the baryonic matter and of radiation are negligible as compared
to the density of dark energy. Anticipating the result of Section \ref%
{Perturbations}, based on a perturbative analysis and CMB temperature power
spectrum, we also include CDM with $\Omega _{CDM}=0.22$.

In the presence of radiation, baryonic matter and CDM components, from among
the equations (\ref{Hhat})-(\ref{sEq}) only (\ref{Hhat}) is changed: 
\begin{eqnarray}
\hat{H}^{2} &=&\frac{\hat{V}}{\sqrt{1-s^{2}}}+\left( \Omega _{b}+\Omega
_{CDM}\right) \left( 1+z\right) ^{3}+\Omega _{rad}\left( 1+z\right) ^{4}\ ,  \label{Hhat2}
\end{eqnarray}%
where $\Omega _{rad}$ is the radiation component (electromagnetic radiation
and massless neutrinos). The symmetry (\ref{sym}) of the system continues to
hold. However since $\hat{H}$ is not proportional to $\sqrt{\Omega _{\Lambda
}}$, in contrast with the pure tachyonic model, the tachyonic field
equations (\ref{yEq})-(\ref{sEq}) are sensitive to $\Omega _{\Lambda }$.
Therefore the initial conditions fixed by $y_{0}$, $s_{0}$ cannot be
represented on a single velocity phase diagram, in general different pairs
of $y_{0}$, $s_{0}$ would generate different $\Omega _{\Lambda }$-s through
Eq. (\ref{Hhat2}), evaluated at $z=0$.

\begin{figure}[th]
\begin{center}
\includegraphics[height=10.0cm,angle=270]{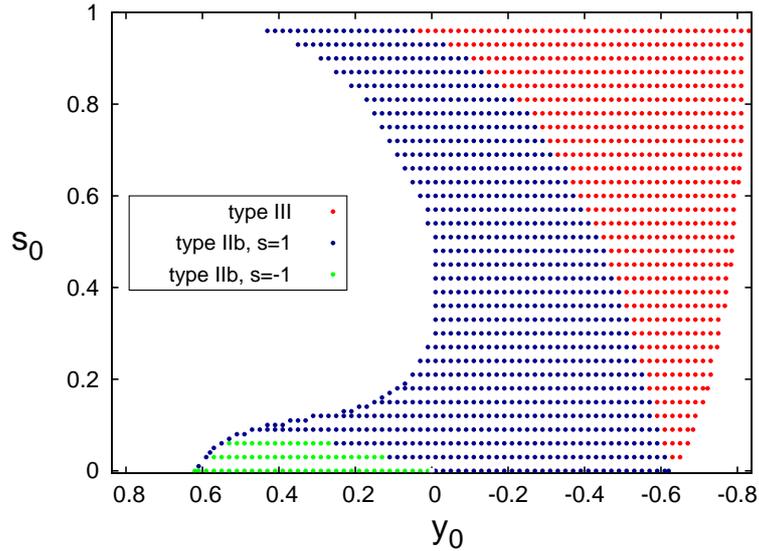} \vskip0.4cm
\end{center}
\caption{The initial conditions at $z=0$ for evolutions of type IIb (blue
and green dots) and III (red dots) in the parameter space ($y_{0}$,$s_{0}$).
The trajectories of type IIb originate from Big Bang type singularities
lying on the lines $s=1$ (blue dots) or $s=-1$ (green dots).}
\label{Fig3}
\end{figure}

On Fig. \ref{Fig3} we represented the initial data for the evolutions of
type IIb and III in the parameter space ($y_{0}$,$s_{0}$). Blue and green
dots denote the set of initial conditions at $z=0$ for the evolutions of
type IIb originating from Big Bang type singularities lying on the lines $%
s=1 $ and $s=-1$, respectively. The red dots represent initial conditions
for the evolutions of type III. Due to the symmetry (\ref{sym}), only the
region $s_{0}\geq 1$ is shown. In the rest of this section we will restrict
these domains by confrontation with various cosmological tests.

\subsection{SNIa, BAO distance ratios, Hubble parameter and CMB acoustic
scale tests}

We will confront the enhanced tachyonic model with both the SNIa and Hubble
parameter data in a similar manner as described in the subsection \ref{Test1}
for the pure tachyonic model. We will also test the model with BAO data,
which determine the ratio:%
\begin{equation}
d_{z}\left( z\right) =\frac{r_{s}\left( z_{drag}\right) }{D_{V}\left(
z\right) }~,  \label{dz}
\end{equation}%
at different redshifts. Here $D_{V}\left( z\right) $ is the volume distance:%
\begin{equation}
D_{V}^{3}\left( z\right) =\frac{zD_{A}^{2}\left( z\right) }{H\left( z\right) 
}~,
\end{equation}%
with comoving angular diameter $D_{A}\left( z\right) $ which can be
expressed by the luminosity distance as%
\begin{equation}
D_{A}\left( z\right) =\frac{d_{L}\left( z\right) }{1+z}~.
\end{equation}%
The quantity $r_{s}$ denotes the sound horizon:%
\begin{equation}
r_{s}\left( z\right) =\int_{z}^{\infty }\frac{dz^{\prime }}{H\sqrt{3\left(
1+R\right) }}~,
\end{equation}%
with $R=3\rho _{b}/4\rho _{\gamma }$, where $\rho _{b}$ and $\rho _{\gamma }$
are the energy densities of the baryons and photons, respectively. The sound
horizon in (\ref{dz}) is evaluated at the baryon drag epoch ($z_{drag}$)
when the baryon velocity perturbations decouple from the photon dipole $%
q_{a}^{\left( \gamma \right) }$. This happens approximately when the baryon
drag optical depth%
\begin{equation}
\tau _{drag}\left( z\right) =\int_{0}^{z}\frac{n_{e}\sigma _{T}}{\left(
1+z^{\prime }\right) HR}dz^{\prime }
\end{equation}%
reaches unity ($\tau _{drag}\left( z_{drag}\right) =1$) \cite{HuSugiyama1996}%
, \cite{CAMB1}. Here $n_{e}$ is the number density of free electrons
(without reionization history) and $\sigma _{T}$ is the Thompson cross
section. The determination of $z_{drag}$ requires to know $n_{e}\left(
z\right) $ from some recombination model. We compute $z_{drag}$ and $%
r_{s}\left( z_{drag}\right) $ numerically from a modified version of the
CAMB code \cite{CAMB1}, \cite{CAMB2}, \cite{CAMB3} in which we implemented
the evolution of the tachyonic universe. For modeling the recombination
history we used the RECFAST subcode \cite{RECFAST}.

Six data on BAO and their inverse covariance matrix $C^{-1}$ applied in the
analysis are given by Table 3 of Ref. \cite{BAOI} and by Eq. (4.3) of Ref. 
\cite{BAOII}, respectively. From the theoretically derived $d_{z}^{th}\left(
z_{i}\right) $ ($i=1,..,6$) and from the observations $d_{z}^{obs}\left(
z_{i}\right) $ a six dimensional vector $\mathbf{X}$ is constructed
containing $d_{z}^{th}\left( z_{i}\right) -d_{z}^{obs}\left( z_{i}\right) $
in its $i$th row. In the cosmological test of the enhanced tachyonic
universe model we computed%
\begin{equation}
\chi _{BAO_{1}}^{2}=\mathbf{X}^{T}C^{-1}\mathbf{X}~,
\end{equation}%
where $T$ denotes the transposed vector. We also include the Baryon
Oscillation Spectroscopic Survey \cite{BOSS} result $d_{z}\left( 0.57\right)
=0.0731\pm 0.0018$ \cite{BOSSdz} by defining%
\begin{equation}
\chi _{BAO}^{2}=\chi _{BAO_{1}}^{2}+\left( \frac{d_{z}^{th}\left(
0.57\right) -0.0731}{0.0018}\right) ^{2}~.
\end{equation}

Before decoupling the acoustic oscillations in a baryon-photon plasma induce
an oscillatory pattern in the CMB temperature. For adiabatic fluctuations,
the $m$th Doppler peak has comoving wave number $k_{m}=m\pi /r_{s}\left(
z_{\ast }\right) $ \cite{HuSugiyama1995}. Here $z_{\ast }$ is the redshift
when the photons decouple from baryons, i.e. when%
\begin{equation}
\tau \left( z\right) =\int_{0}^{z}\frac{n_{e}\sigma _{T}}{\left( 1+z^{\prime
}\right) H}dz^{\prime }
\end{equation}%
reaches unity ($\tau \left( z_{\ast }\right) =1$). The location of the first
peak of the CMB temperature spectrum in multipole space is%
\begin{equation}
l_{A}\approx \frac{\pi D_{A}\left( z_{\ast }\right) }{r_{s}\left( z_{\ast
}\right) }~.
\end{equation}%
We test the tachyonic universe model with the CMB acoustic scale $%
l_{A}^{obs}=301.65\pm 0.18~$\cite{PlancklA} by computing the following $\chi
^{2}$ value:%
\begin{equation}
\chi _{CMB}^{2}=\left( \frac{l_{A}^{th}-301.65}{0.18}\right) ^{2}~,
\end{equation}%
where $l_{A}^{th}$ is derived numerically by the modified CAMB code.


\begin{figure}[th]
\includegraphics[height=7.2cm,angle=270]{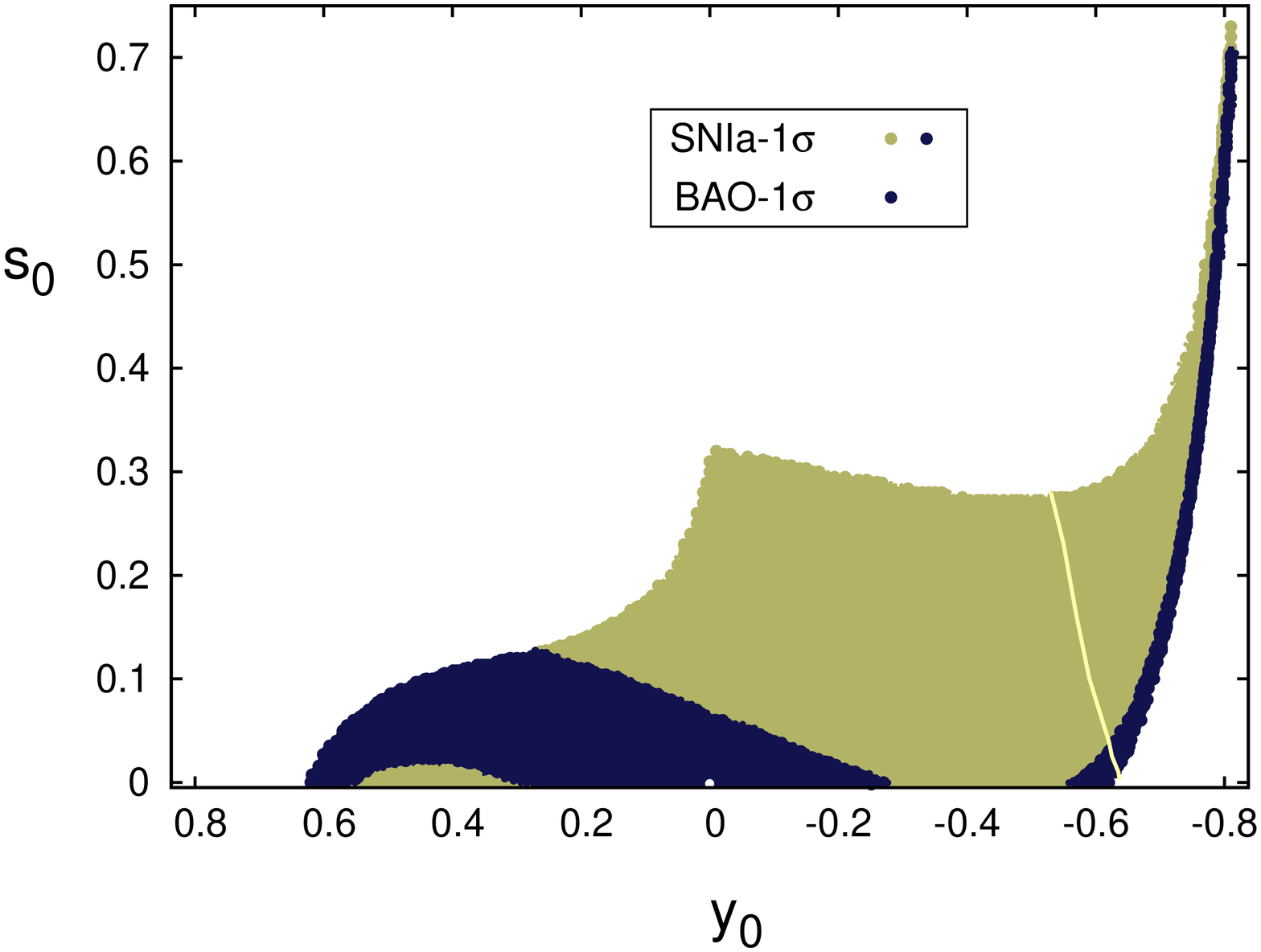} \hskip0.4cm%
\includegraphics[height=7.2cm,angle=270]{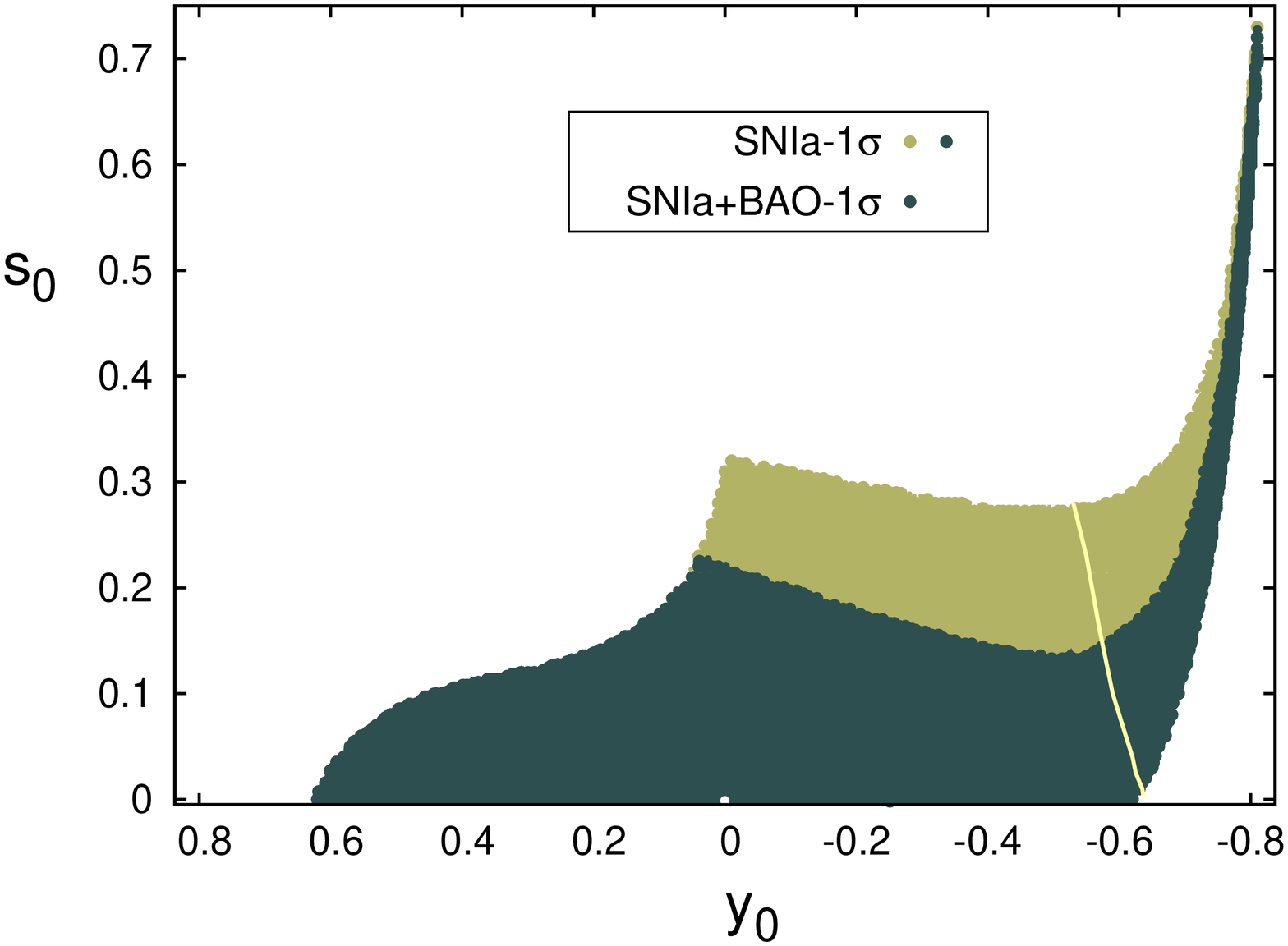} \vskip0.4cm %
\includegraphics[height=7.2cm,angle=270]{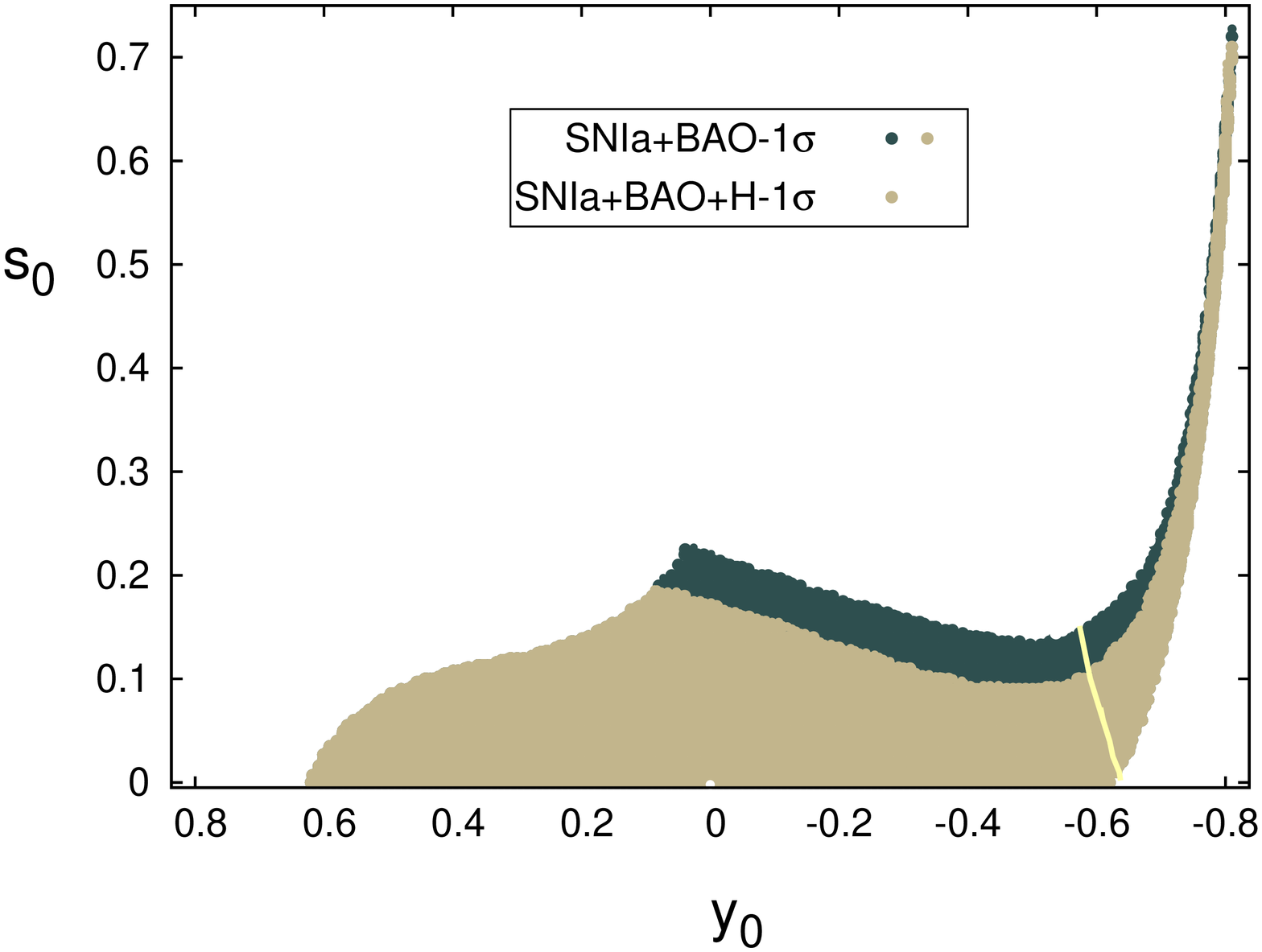} \hskip0.4cm%
\includegraphics[height=7.2cm,angle=270]{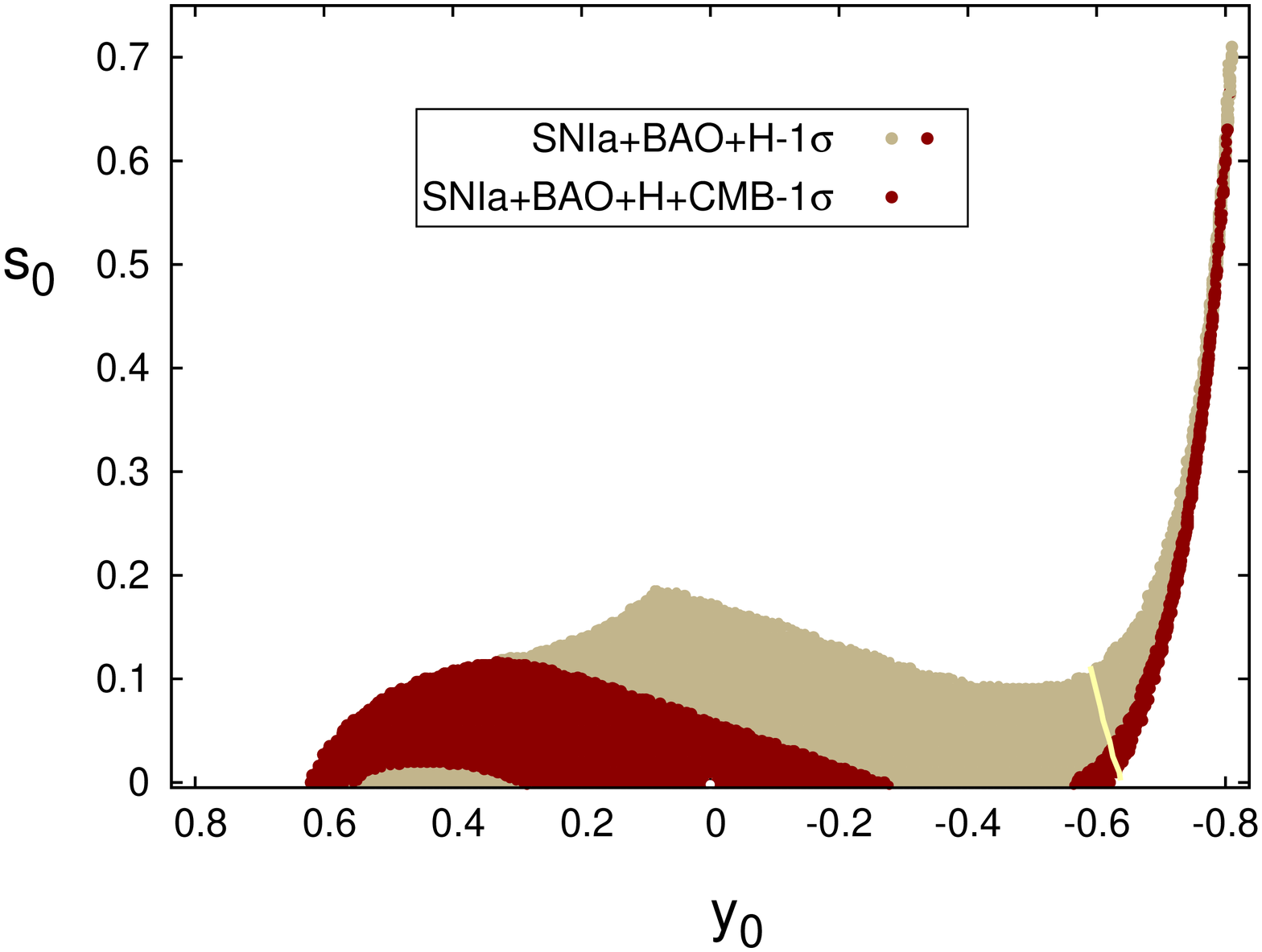} \vskip0.4cm
\caption{The initial values ($y_{0}$,$s_{0}$) of the tachyonic parameter
space constrained by cosmological tests. Upper left panel: the 1$\protect%
\sigma $\ confidence level regions selected by the SNIa (mint green) and by
BAO (navy blue), the latter masking out some of the more extended SNIa 1$%
\protect\sigma $\ region. Upper right: the inclusion of BAO reduces the 1$%
\protect\sigma $\ confidence level region of the SNIa (mint green) to the
one of the SNIa+BAO (myrtle green). Lower left: the inclusion of Hubble
parameter data set further reduces the 1$\protect\sigma $\ confidence level%
\textbf{\ }region of the SNIa+BAO (myrtle green) to the one of SNIa+BAO+H
(ecru). Lower right: the inclusion of CMB acoustic scale reduces once again
the 1$\protect\sigma $\ confidence level region of the SNIa+BAO+H (ecru) to
the one of SNIa+BAO+H+CMB (crimson red). All 1$\protect\sigma $ parameter
regions are divided into a left region (representing type IIb trajectories)
and a right region (representing type III trajectories) by a yellow line.
The $\Lambda $CDM attractor in the origin is represented by a white
dot.}
\label{Fig4}
\end{figure}


On Fig. \ref{Fig4} the regions of parameter space ($y_{0}$,$s_{0}$) are
shown in which the tachyonic universe model fits with the above data sets at
1$\sigma $ confidence level. On each panel the yellow curve separates the
regions of initial conditions for trajectories of types IIb (on the left)
and III (on the right). On the upper left panel the colored regions
represent the fitting of the model at 1$\sigma $ confidence level with SNIa
(both the mint green and navy blue regions) and BAO (the navy blue region)
data sets. The BAO distance ratio test leads to a much stringent restriction
of the parameter space than the SNIa test. However we note that the BAO data
set consists of much less data (7) than the SNIa set (580), and the $\chi
^{2}$-test works better for a larger number of data points. On the
subsequent three panels we show how the inclusion of each of the BAO
distance ratios, Hubble parameter-redshift relation and CMB acoustic scale
cosmological test successively restrict the 1$\sigma $\ region of the SN1a
test.

In particular, on the upper right panel the subset of the SNIa-1$\sigma $
domain which fits to the SNIa+BAO data set at 1$\sigma $ confidence level is
shown in myrtle green. This combined test was performed by computing $\chi
_{SNIa+BAO}^{2}=\chi _{SNIa}^{2}+\chi _{BAO}^{2}$. The SNIa+BAO test is less
restrictive as compared with the test of BAO only, since the BAO-1$\sigma $
domain is included into the SNIa-1$\sigma $ domain and the critical $\chi
^{2}$ belonging to the 1$\sigma $ confidence level increases with the amount
of data. Since the SNIa test is based on a significantly larger amount of
data than the number of BAO distance ratios, the fitting with SNIa data
dominates the combined test. On the lower left panel the 1$\sigma $\
confidence region resulted from the $\chi _{SNIa+BAO+H}^{2}=\chi
_{SNIa}^{2}+\chi _{BAO}^{2}+\chi _{H}^{2}$ test with the Hubble parameter
data set also included\ is shown in ecru. Finally, we add the CMB acoustic
scale to the test by calculating $\chi _{SNIa+BAO+H+CMB}^{2}=\chi
_{SNIa}^{2}+\chi _{BAO}^{2}+\chi _{H}^{2}+\chi _{CMB}^{2}$ which further
restricts the domain of the parameter space which fits at 1$\sigma $
confidence level, shown in crimson red on the lower right panel. This is
quite similar in shape and size to the BAO-1$\sigma $ domain.

The tale of Fig. \ref{Fig4} is that there are trajectories of both types IIb
and III which survive the various combined tests. By comparing Figs. \ref%
{Fig3} and \ref{Fig4} we conclude that a relatively larger subset of
evolutions of type IIb fit the data at 1$\sigma $ confidence level as
compared to the trajectories of type III. In this sense, from the two
possible future scenarios, the evolutions ending in de Sitter attractor are
more likely.

\section{Cosmic microwave background in type IIb and III tachyonic universe
models \label{Perturbations}}

In this section we perturb the flat Friedmann universe in order to derive
the CMB temperature spectrum. As usual the perturbations of the Friedmann
universe are classified into scalar, vector and tensor types. Here we
investigate only the scalar type perturbations by implementing the source
terms due to perturbations of the tachyonic scalar field and their evolution
equations into the freely available CAMB code.

The perturbation equations in the CAMB code were derived in the framework of
3+1 covariant formalism \cite{CAMB1},\cite{CAMB2}, \cite{form}, \cite{exp}
in which the space-time metric $g_{ab}$ is split in the form $%
g_{ab}=u_{a}u_{b}+h_{ab}$, with $u^{a}u_{a}=1$ and $u^{a}h_{ab}=0$. Here $%
h_{ab}$ is the projection tensor into the rest space of an observer moving
with 4-velocity $u^{a}$. In the Friedmann space-time a convenient choice is $%
u_{a}=\left( dt\right) _{a}$ which is the comoving system with the matter
flow. In the perturbed Friedmann space-time there are infinite possible
choices for $u_{a}$ which coincide with $\left( dt\right) _{a}$ in the
absence of perturbations. In the CAMB code for the scalar type perturbation
the frame (i.e. $u_{a}$) is defined by $A_{b}=u^{b}\nabla _{b}u_{a}=0$,
where $\nabla _{a}$ is the covariant derivative. The scalar type velocity
perturbations of the CDM vanish in this so called CDM frame. This
description of the perturbations corresponds to the choice of synchronous
gauge in a metric based perturbation formalism \cite{Ma}.

The tachyonic scalar field interacts with the other matter components only
gravitationally. Therefore the contributions arising from the tachyonic
field to the equations governing the perturbations of other matter
components appear exclusively through the changes induced in the space-time
curvature. In other words no particle scattering processes between the
tachyonic and other matter components are allowed. In the 3+1 covariant
formalism the energy-momentum tensor of the perturbed matter is described in
terms of $D_{a}\rho $, $D_{a}p$, $q$ and $\pi _{ab}$. After we will enlist
the contributions of the tachyonic field, we will derive the evolution
equations governing the perturbations.

The tachyonic energy-momentum tensor, as arising from the variation of its
action with respect to the metric, and applying the decomposition (\ref%
{EMTdec}) leads to%
\[
\rho ^{\left( T\right) }=V\sqrt{1-2X}+\frac{V\dot{T}^{2}}{\sqrt{1-2X}}~, 
\]%
\[
p^{\left( T\right) }=-\frac{V\left( D^{a}T\right) \left( D_{a}T\right) }{3%
\sqrt{1-2X}}-V\sqrt{1-2X}~, 
\]%
\[
q_{a}^{\left( T\right) }=\frac{V\dot{T}D_{a}T}{\sqrt{1-2X}}~, 
\]%
\begin{equation}
\pi _{ab}^{\left( T\right) }=\frac{V\left( D_{\langle a}T\right) \left(
D_{b\rangle }T\right) }{\sqrt{1-2X}}~.  \label{TachFluid}
\end{equation}%
where%
\begin{equation}
X=\frac{1}{2}\left( \nabla ^{a}T\right) \left( \nabla _{a}T\right) \mathcal{~%
},
\end{equation}%
$D_{a}$ is the covariant derivative on the 3-space with metric $h_{ab}$ ($%
D_{a}T=h_{a}^{b}\nabla _{b}T$) and the dot denotes: $\dot{T}=u^{a}\nabla
_{a}T$ (in the absence of perturbations this coincides with the time
derivative employed at the background level). The angular bracket $\langle $ 
$\rangle $ on abstract indices denotes the trace free part of a symmetrized
tensor projected in all indices with the metric $h_{ab}$.

The 3+1 covariant equations governing the perturbations at first order
contain the background values of $\rho ^{\left( T\right) }$ and $p^{\left(
T\right) }$, their spacelike derivatives ($D_{a}\rho ^{\left( T\right) }$
and $D_{a}p^{\left( T\right) }$) and also the quantities $q_{a}^{\left(
T\right) }$ and $\pi _{ab}^{\left( T\right) }$. At first order in the
perturbations we find%
\[
D_{a}\rho ^{\left( T\right) }=\frac{Vs}{\left( 1-s^{2}\right) ^{3/2}}D_{a}%
\dot{T}+\frac{V_{,T}}{\sqrt{1-s^{2}}}D_{a}T~, 
\]%
\[
D_{a}p^{\left( T\right) }=\frac{Vs}{\sqrt{1-s^{2}}}D_{a}\dot{T}-\sqrt{1-s^{2}%
}V_{,T}D_{a}T~, 
\]%
\begin{equation}
q_{a}^{\left( T\right) }=\frac{Vs}{\sqrt{1-s^{2}}}D_{a}T~,~~\pi
_{ab}^{\left( T\right) }=0~.
\end{equation}%
Here $s$ denotes the background value of $\dot{T}$. From the three
nonvanishing quantities describing the perturbed field only two are
independent since the pressure gradient can be expressed as%
\begin{equation}
D_{a}p^{\left( T\right) }=\left( 1-s^{2}\right) \left[ D_{a}\rho ^{\left(
T\right) }-\frac{2V_{,T}}{Vs}q_{a}^{\left( T\right) }\right] ~.
\label{press}
\end{equation}

In what follows we apply a harmonic expansion in order to derive ordinary
differential equations for the variables characterizing the perturbation in
Friedmann space-time. The scalar harmonics are the eigenfunctions of the
spatial Laplacian:

\begin{equation}
D^{2}Q^{S\left( k\right) }=\frac{k^{2}}{a^{2}}Q^{S\left( k\right) }\ ,
\end{equation}%
with $\dot{Q}^{S\left( k\right) }=0$ at zeroth order. From $Q^{S\left(
k\right) }$ we construct the following projected vector and symmetric
trace-free tensor:%
\begin{equation}
Q_{a}^{S\left( k\right) }=\frac{a}{k}D_{a}Q^{S\left( k\right) }\ ,\ \
Q_{ab}^{S\left( k\right) }=\frac{a^{2}}{k^{2}}D_{\langle a}D_{b\rangle
}Q^{S\left( k\right) }\ .
\end{equation}%
The 3-vectors and symmetric trace-free 3-tensors arising from scalar
perturbations can be expanded in terms of $Q_{a}^{S\left( k\right) }$ and $%
Q_{ab}^{S\left( k\right) }$, respectively \cite{CAMB1}, \cite{exp}. The
harmonic expansions for the tachyonic field variables are 
\begin{equation}
\delta _{a}^{\left( T\right) }\equiv \frac{a}{\rho ^{\left( T\right) }}%
D_{a}\rho ^{\left( T\right) }=\sum_{k}k\delta _{k}^{\left( T\right)
}Q_{a}^{S\left( k\right) }\ ,
\end{equation}%
\begin{equation}
D_{a}p^{\left( T\right) }=\rho ^{\left( T\right) }\sum_{k}\frac{k}{a}%
p_{k}^{\left( T\right) }Q_{a}^{S\left( k\right) }\ ,
\end{equation}%
\begin{equation}
q_{a}^{\left( T\right) }=\left( \rho ^{\left( T\right) }+p^{\left( T\right)
}\right) \sum_{k}v_{k}^{\left( T\right) }Q_{a}^{S\left( k\right) }\ .
\end{equation}%
Defining the expansion of $D_{a}T$ as%
\begin{equation}
D_{a}T=\sum_{k}\frac{k}{a}T_{k}Q_{a}^{S\left( k\right) }\ ,
\end{equation}%
we find%
\begin{equation}
v_{k}^{\left( T\right) }=\frac{k}{as}T_{k}~,  \label{vdef}
\end{equation}%
\begin{equation}
\delta _{k}^{\left( T\right) }=\frac{s}{1-s^{2}}\dot{T}_{k}+\frac{V_{,T}}{V}%
T_{k}~,
\end{equation}%
where we have used the commutation relation: $aD_{a}\dot{T}=\left(
aD_{a}T\right) ^{\cdot }$ which is valid in the CDM frame. The harmonic
coefficient ($p_{k}^{\left( T\right) }$) arises from harmonic decomposition
of Eq. (\ref{press}): 
\begin{equation}
p_{k}^{\left( T\right) }=\left( 1-s^{2}\right) \left[ \delta _{k}^{\left(
T\right) }-\frac{2V_{,T}}{V}\frac{as}{k}v_{k}^{\left( T\right) }\right] ~.
\end{equation}

The equations of motion for $v_{k}^{\left( T\right) }$ and $\delta
_{k}^{\left( T\right) }$ follow from the divergenceless condition of the
energy-momentum tensor for tachyonic field ($\nabla ^{a}T_{ab}^{\left(
T\right) }=0$) and a harmonic expansion. Taking the 3-gradient of the
projection $u^{b}\nabla ^{a}T_{ab}^{\left( T\right) }=0$ gives the evolution
equation for $\delta _{a}^{\left( T\right) }$, while the equation governing $%
q_{a}^{\left( T\right) }$ emerges from the projection $h_{c}^{b}\nabla
^{a}T_{ab}^{\left( T\right) }=0$. Then the harmonic expansion generates the
equations of motion for $v_{k}^{\left( T\right) }$ and $\delta _{k}^{\left(
T\right) }$.

At this point it is worth to introduce the variable%
\begin{equation}
\mathcal{X}_{k}^{\left( T\right) }=k\delta _{k}^{\left( T\right)
}+3aHs^{2}v_{k}^{\left( T\right) }=\frac{ks^{2}}{1-s^{2}}\left( \frac{T_{k}}{%
s}\right) ^{\cdot }  \label{CallXvar}
\end{equation}%
replacing $\delta _{k}^{\left( T\right) }$, as the evolution equation for $%
\mathcal{X}_{k}^{\left( T\right) }$ becomes simpler. For the second equality
of (\ref{CallXvar}) we have employed Eq. (\ref{KG}). Note that the original
Fourier components $\left( T_{k},\dot{T}_{k}\right) $ of the velocity
phase-space variables originally replaced by $\left( v_{k}^{\left( T\right)
},\delta _{k}^{\left( T\right) }\right) $ are changed into $\left(
v_{k}^{\left( T\right) },\chi _{k}^{\left( T\right) }\right) $.

The equations of motion for the tachyonic scalar field perturbations in the
Fourier space read%
\begin{equation}
v_{k}^{\left( T\right) \prime }=-\mathcal{H}v_{k}^{\left( T\right) }+\frac{%
1-s^{2}}{s^{2}}\mathcal{X}_{k}^{\left( T\right) }~,  \label{PEq2}
\end{equation}%
\begin{eqnarray}
\mathcal{X}_{k}^{\left( T\right) \prime } &=&-3\left( 1-s^{2}\right) 
\mathcal{HX}_{k}^{\left( T\right) }-k^{2}s^{2}\left( \mathcal{Z}%
_{k}+v_{k}^{\left( T\right) }\right)   \nonumber \\
&&+3s^{2}\left( \mathcal{H}^{\prime }-\mathcal{H}^{2}\right) v_{k}^{\left(
T\right) }~,  \label{PEq1}
\end{eqnarray}%
where $\mathcal{H}=a^{\prime }/a$ and the prime denotes the derivative with
respect to the conformal time $\eta $ introduced as $ad\eta =dt$. The
variable $\mathcal{Z}_{k}$ determines the harmonic coefficient of the
comoving spatial gradient of the expansion $\Theta =D^{a}u_{a}$ ($=3H$ in
the unperturbed Friedmann space-time) as%
\begin{equation}
\mathcal{Z}_{a}\equiv aD_{a}\Theta =\sum_{k}\frac{k^{2}}{a}\mathcal{Z}%
_{k}Q_{a}^{S\left( k\right) }~.
\end{equation}

By virtue of the definitions (\ref{vdef}) and (\ref{CallXvar}) the equation (%
\ref{PEq2}) is identically satisfied, while Eq. (\ref{PEq1}) gives the
following second order equation for $T_{k}$:%
\begin{equation}
\left( \frac{\dot{T}_{k}}{1-s^{2}}\right) ^{\cdot }=-3H\dot{T}_{k}-\left[
\left( \frac{V_{,T}}{V}\right) _{,T}+\frac{k^{2}}{a^{2}}\right] aT_{k}-ks%
\mathcal{Z}_{k}~.
\end{equation}%
This equation can also be derived directly from the action for the tachyonic
scalar field, at linear older in the perturbations.


\begin{figure}[th]
\includegraphics[height=7.2cm,angle=270]{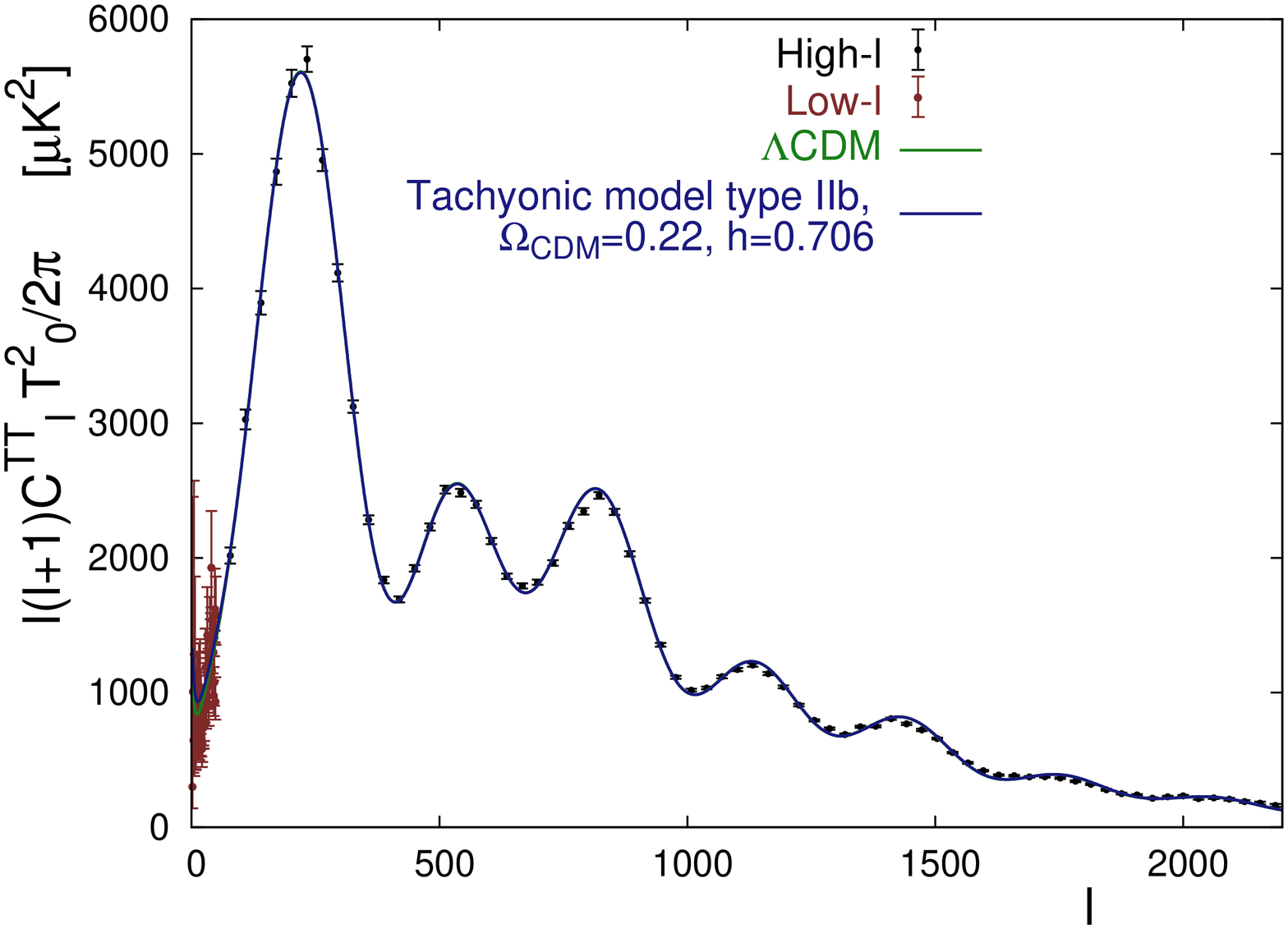} %
\hskip0.4cm%
\includegraphics[height=7.2cm,angle=270]{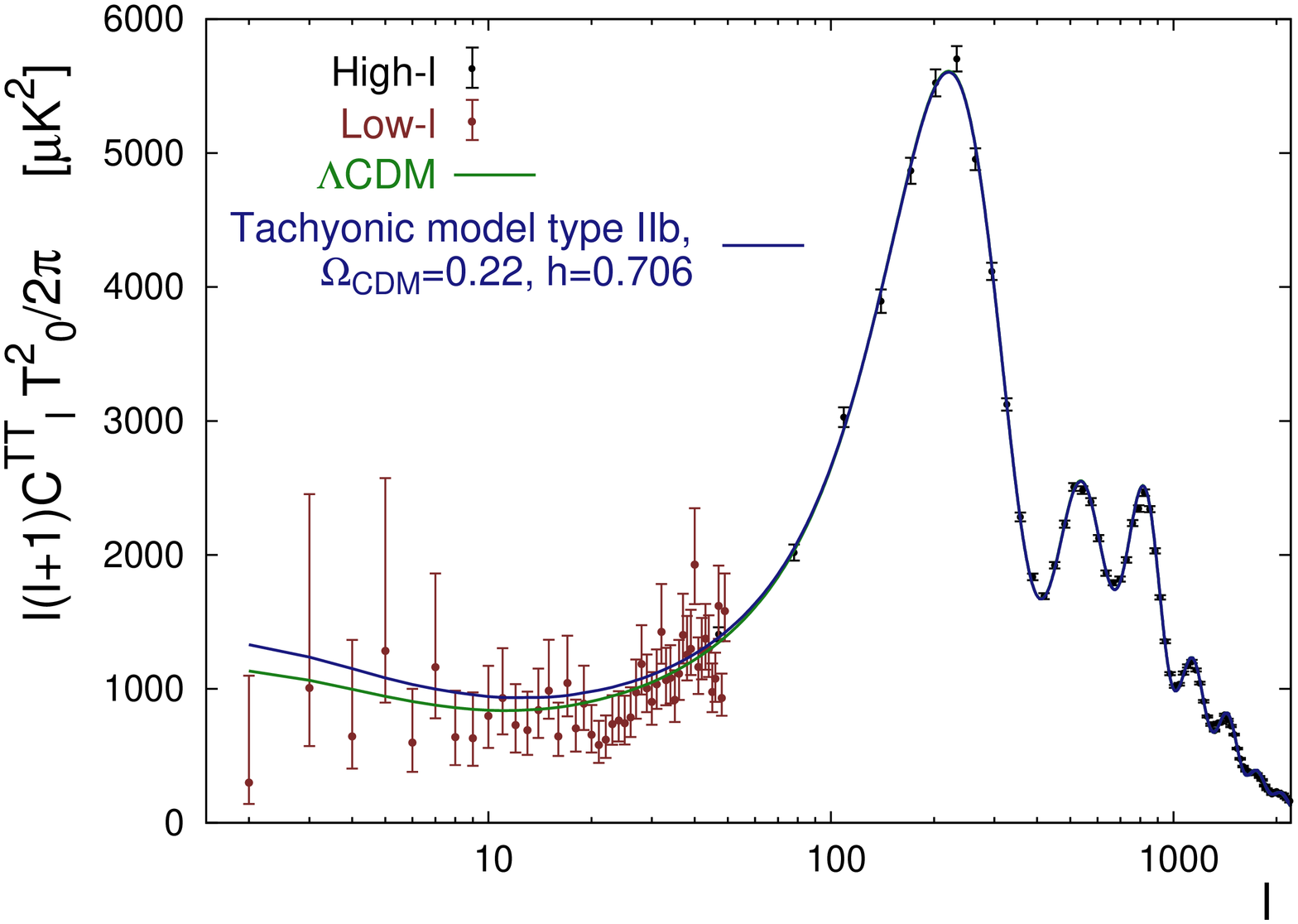} %
\vskip0.4cm %
\includegraphics[height=7.2cm,angle=270]{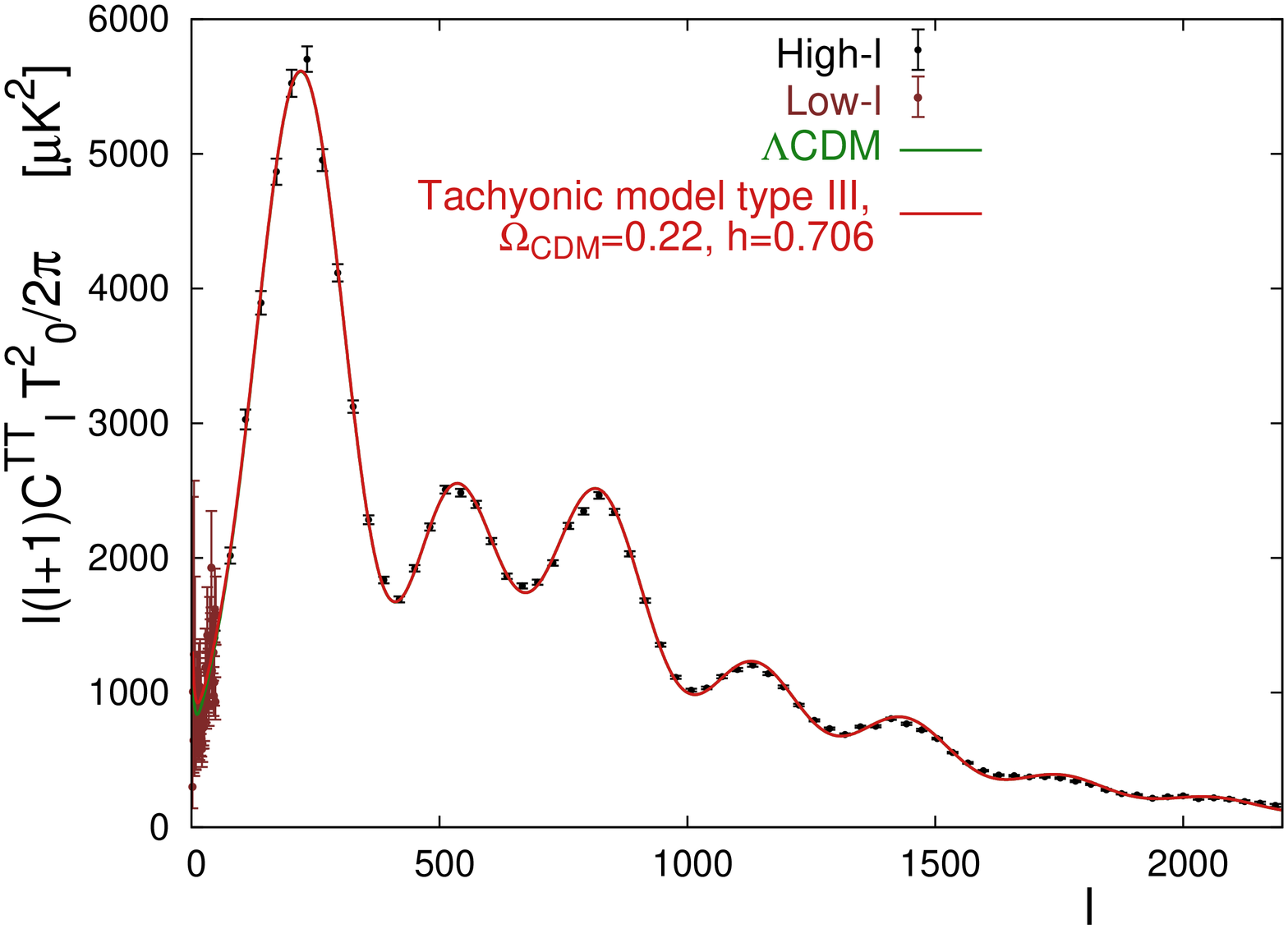} %
\hskip0.4cm%
\includegraphics[height=7.2cm,angle=270]{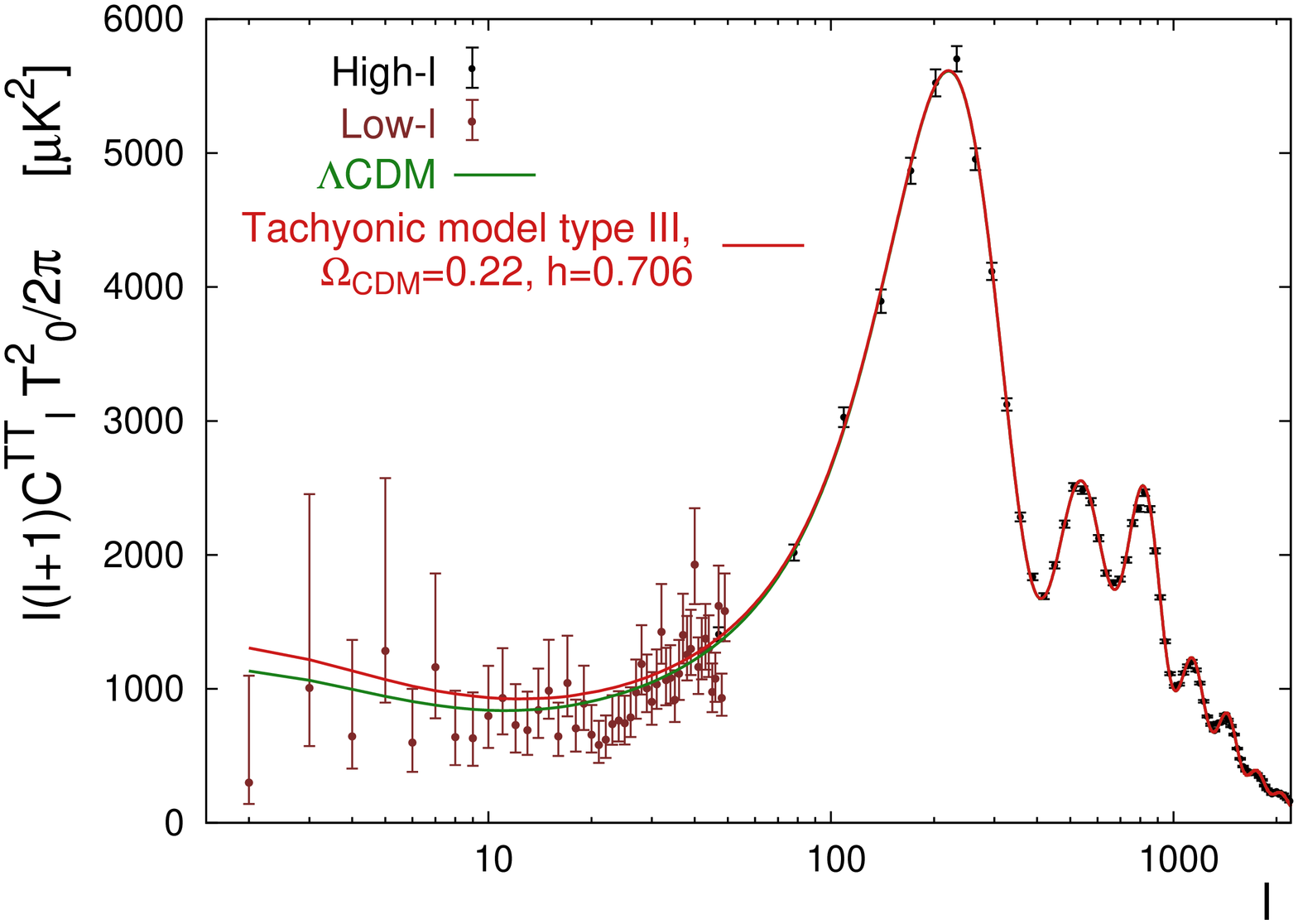} %
\vskip0.4cm
\caption{The CMB temperature power spectrums is shown for evolutions of type
IIb (upper row, blue line) and III (lower row, red line) on linear multipole 
$l$ scale (left columns) and logarithmic scale (right columns). The green
line represents the best fit $\Lambda $CDM model (its parameters are given
by the last column in Table 5 of \protect\cite{PlanckXVI}). Data points with
error bars are shown with brown for Low$-l$ and black for High$-l$
multipoles.}
\label{Fig5}
\end{figure}


At high redshift $s^{2}\approx 1$, the tachyonic scalar field behaves as
CDM, therefore we can choose the same adiabatic initial conditions as for
CDM at $z\approx 10^{9}$. The parameters such as the Thompson scattering
optical depth due to reionization $\tau $, the scalar spectral index $n_{s}$%
, the power of the primordial scalar curvature perturbation $A_{s}$ are
taken from the $\Lambda $CDM model (the last column in Table 5 of \cite%
{PlanckXVI}). Other parameters affecting the CMB temperature power spectrum
are $H_{0}$, $\Omega _{b}$ and$\ \Omega _{CDM}$. From among these we already
fixed the first two.

By varying the last one over the range $\Omega _{CDM}\in \left( 0,0.3\right) 
$ we found the most reasonable CMB\ temperature power spectrum for the value
of $\Omega _{CDM}=0.22$ for either type of evolutions. On Fig. \ref{Fig5} we
represented the best fit CMB temperature power spectrum for the evolutions
of type IIb (upper row, blue line) and III (lower row, red line) on linear
multipole $l$ scale (left columns) and logarithmic scale (right columns).
For comparison the green line representing the best fit $\Lambda $CDM model
is also shown. The data set and their error bars are given in brown for Low$%
-l$ and black for High$-l$ multipoles. The CMB spectrum in the tachyonic
universe model fits the Planck data as well as the standard $\Lambda $CDM
model at high multipoles. At low multipoles the power is somewhat higher
than in the case of the $\Lambda $CDM model.

\section{Concluding Remarks \label{CR}}

We investigated a tachyonic scalar field model governed by a trigonometric
potential, which exhibits a rich variety of future evolutions, depending on
the initial data and the actual value of the model parameters. For a
positive model parameter $k=0.44$ chosen for this paper, all possible
evolutions originate in Big Bang type singularities, while they end either
in a de Sitter exponential expansion (trajectories of type I and II) or into
a sudden future singularity (types III, IV and V).

Previous confrontation with SNIa data confirmed that despite so different,
both future scenarios are compatible with the hypothesis of dark energy in
the form of this tachyonic scalar field. It this paper we clarified which
types are allowed by observations. Type V evolutions being confined to the
positive pressure regions, never achieve accelerated expansion, hence they
are excluded. A careful analysis of the model of a Friedmann universe filled
with the tachyonic scalar field identified that all evolutions compatible
with the Union 2.1 SNIa data are of types I, II and III only, those of type
IV running outside the 1$\sigma $ contours of the SNIa test.\ Furthermore,
the Hubble parameter data set test revealed that while at 1$\sigma $\
confidence level the trajectories of types II and III remain compatible, the
trajectories of type I are disruled.

On the other hand, the negative pressure of the tachyonic scalar field
decreases fast in magnitude during the backward evolution in time of the
trajectories of types I-III. The past behavior of the evolutions of type II
was not well understood before. Indeed, it was unclear whether the
separatrix between the evolutions of type II and III reaches the corner
point P (P'), in other words whether there are evolutions of type II with
eternal negative pressure. Our present analysis has elucidated that such
evolutions are possible, leading to a further subclassification of the
trajectories of type II into the subtype IIa (trajectories born from a Big
Bang with positive pressure, evolving superluminally, then passing through
the corner point P (P') and becoming dark energy with negative pressure
nowadays) and IIb (born from a Big Bang with negative pressure). A new
separatrix between these subtypes IIa and IIb, starting from the point P
(P') was added to the velocity phase diagram, Fig. \ref{Fig1}.

Both evolutions of types IIb and III become dust-like in the past,
suggesting a degeneration into dark matter of the dark energy scalar field.
By contrast the trajectories of type I and some of those of IIa exhibit a
large build-up of pressure in the distant past, disruling them as viable
cosmological models explaining BBN. Another immediate argument for
invalidating the evolutions of type I and all evolutions of type IIa is a
negative speed of sound squared in the regions with positive pressure, which
in general leads to instabilities in the evolution of perturbations. Such
instabilities would not allow the universe to reach its present state.
(Similar instabilities could drastically affect the future evolution of the
trajectories of type III, such that they would be hampered to reach the soft
singularities. Such a discussion falls beyond the scope of the present
paper.) As the evolutions IIb and III are both compatible with the SNIa and
Hubble parameter data sets and they do not suffer from instabilities, a more
thorough analysis of these evolutions has been performed in the rest of the
paper.

In order to get a viable cosmological model, radiation, baryonic matter and
CDM constituents were included into the model, complementing the dominant
tachyonic scalar field. Then the evolutions of type IIb and III of this
4-component model were confronted with a series of cosmological tests at the
background level, including the supernova type Ia Union 2.1 data, BAO
distance ratios, the Hubble parameter data and the CMB acoustic scale. We
identified the evolutions of both types, which at 1$\sigma $ confidence
level survive these cosmological tests.

\begin{figure}[bh]
\begin{center}
\includegraphics[height=12.0cm,angle=270]{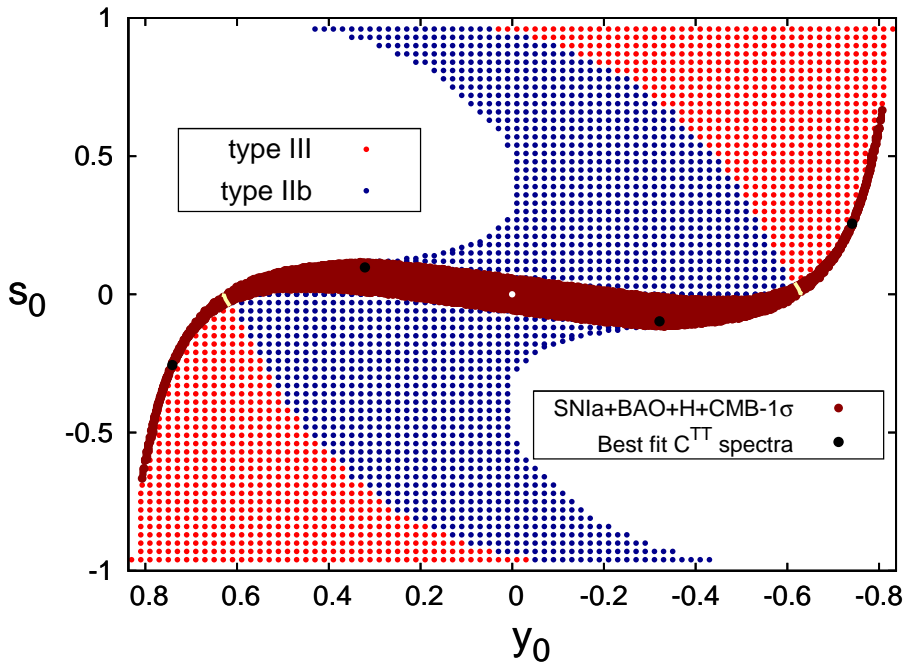} \vskip0.4cm
\end{center}
\caption{The initial conditions (at $z=0$) in the parameter space ($y_{0}$,$%
s_{0}$) for the evolutions of type IIb (blue dots) and III (red dots), as
restricted by the combined cosmological test of the SNIa data, BAO distance
ratios, Hubble parameter data and CMB acoustic scale, evaluated for $\Omega
_{CDM}=0.22$. The initial data where the CMB temperature power spectrum
showed the best fit for each type of evolution is indicated by black
bullets. The $\Lambda $CDM attractor in the origin is represented by a white
dot. Although the points ($y_{0}$,$s_{0}$) and ($-y_{0}$,$-s_{0}$) represent
identical tachyonic trajectories, both are shown in order to obtain a
continuos parameter domain (crimson red) compatible with the combined test
at 1$\protect\sigma $ confidence level.}
\label{Fig6}
\end{figure}

A further test of the model evolutions was performed at perturbative level.
There we derived the CMB temperature power spectrum and found the best
agreement with the Planck data for a CDM component with $\Omega _{CDM}=0.22$%
, less than in the $\Lambda $CDM model. The difference in the amount of CDM
is explained by the dust-like behaviour of the tachyonic scalar field. The
fit of the spectrum with the data was similar to the $\Lambda $CDM model at
high multipoles, but the power remained slightly overestimated at low
multipoles, for both types of evolutions. There, however, the fit of the $%
\Lambda $CDM model is also less satisfactory than for the high multipoles,
and any future improvement on the $\Lambda $CDM model to address this could
also improve the fit of the tachyonic model.

In standard cosmology the SNIa test, BAO distance ratios and the location of
the first peak of the CDM temperature power spectrum generate transverse 1$%
\sigma $ contours. In our analysis we have assumed a flat Friedmann
background and the CDM temperature power spectrum selected $\Omega
_{CDM}=0.22$, hence the rest of the tests could be used to restrict the
tachyonic parameters. The result was presented on Fig. \ref{Fig4}: the 1$%
\sigma $\ domain of the SNIa test was severely restricted by BAO distance
ratios, this was successively further reduced by the Hubble parameter data
set and finally by the CMB acoustic scale. The subset of evolutions of types
IIb and III compatible at 1$\sigma $ confidence level with these combined
tests as compared to the full set of evolutions of types IIb and III are
shown on Fig. \ref{Fig6}. A relatively larger subset of the type IIb
evolutions (towards the de Sitter attractor) survive the combined tests, as
compared to the similar evolutions of type III (converging to a future soft
singularity).

In summary we found that a tachyonic scalar field universe enhanced with
radiation, baryonic matter and CDM constituents could well harmonize with
the enlisted observations of our physical universe and the parameter space
of the model compatible with these tests allows for two types of evolutions.
These run similarly in the past, both being born from a Big Bang, with a
subsequent dust-like evolution of the scalar field, achieving scalar field
dark energy driven acceleration at present, but diverging in their future
either into the de Sitter type expansion (type IIb) or by contrast, reaching
a positive pressure regime, leading to a sudden future singularity (type
III). How seriously the latter evolutions would be hampered by instabilities
after crossing the positive pressure divide (the cornerstone P of the phase
diagram) remains a question for future analysis.

\section*{Acknowledgements}

We are grateful for various interactions and discussions on the subject to
Alexander Kamenshchik, Shinji Tsujikawa, Ryotaro Kase and Arman Shafieloo
and acknowledge the helpful suggestions of the referee. The research of both
ZK and L\'{A}G was supported by the European Union and the State of Hungary,
co-financed by the European Social Fund in the framework of T\'{A}MOP 4.2.4.
A/2-11-1-2012-0001 'National Excellence Program'.

\end{document}